\documentclass[aps,prd,twocolumn,10pt,superscriptaddress,nofootinbib,preprintnumbers]{revtex4-2}
\usepackage{amssymb,amsmath,nicefrac}
\usepackage{graphicx}
\usepackage{orcidlink}
\usepackage[indent]{parskip}
\usepackage{microtype}
\usepackage[normalem]{ulem}
\usepackage{hyperref}
\hypersetup{colorlinks,allcolors=[rgb]{0 0.5 1}}

\newcommand{\brac}[2]{{\left( \frac{#1}{#2} \right)}}
\newcommand{\be}{\begin{eqnarray}}
\newcommand{\ee}{\end{eqnarray}}
\newcommand{\beq}{\begin{equation}}
\newcommand{\eeq}{\end{equation}}

\newcommand{\BH}{{\mathrm{BH}}}

\newcommand{\DM}{{\mathrm{DM}}}
\newcommand{\cl}{{\mathrm{cl}}}
\newcommand{\mrg}{{\mathrm{mrg}}}
\newcommand{\form}{{\mathrm{form}}}
\newcommand{\col}{{\mathrm{col}}}
\newcommand{\relic}{{\mathrm{relic}}}
\newcommand{\ev}{{\mathrm{ev}}}

\newcommand{\DNeff}{{\Delta N_\mathrm{eff}}}
\newcommand{\mavg}{{\mkern 2.5mu \overline{\mkern-2.5mu m \mkern-2mu} \mkern 2mu}}

\newcommand{\Eq}[1]{Eq.~(\ref{#1})}

\newcommand{\colorindicator}[1]{{\textcolor[HTML]{#1}{\scalebox{0.8}{$\blacksquare$}}}}

\begin{document}

\preprint{\tt FERMILAB-PUB-24-0133-T}

\title{Clustering and Runaway Merging in a Primordial Black Hole Dominated Universe}

\author{Ian Holst\,\orcidlink{0000-0003-4256-3680}}
\email{holst@uchicago.edu}
\affiliation{Department of Astronomy and Astrophysics, University of Chicago, Chicago, IL 60637}
\affiliation{Kavli Institute for Cosmological Physics, University of Chicago, Chicago, IL 60637}

\author{Gordan Krnjaic\,\orcidlink{0000-0001-7420-9577}}
\email{krnjaicg@fnal.gov}
\affiliation{Department of Astronomy and Astrophysics, University of Chicago, Chicago, IL 60637}
\affiliation{Kavli Institute for Cosmological Physics, University of Chicago, Chicago, IL 60637}
\affiliation{Theoretical Physics Department, Fermi National Accelerator Laboratory, Batavia, Illinois 60510}

\author{Huangyu Xiao\,\orcidlink{0000-0003-2485-5700}}
\email{huangyu@fnal.gov}
\affiliation{Kavli Institute for Cosmological Physics, University of Chicago, Chicago, IL 60637}
\affiliation{Theoretical Physics Department, Fermi National Accelerator Laboratory, Batavia, Illinois 60510}

\date{\today}

\begin{abstract}
If primordial black holes (PBH) are present in the early universe, their contribution to the energy budget grows relative to that of radiation and generically becomes dominant unless the initial abundance is exponentially small. This black hole domination scenario is largely unconstrained for PBHs with masses $\lesssim 10^9\,\mathrm{g}$, which evaporate prior to Big Bang nucleosynthesis. However, if the era of PBH domination is sufficiently long, the PBHs form clusters and can merge appreciably within these objects. We calculate the population statistics of these clusters within the Press-Schechter formalism and find that, for a wide range of PBH masses and Hubble rates at the onset of PBH domination, the mergers within PBH clusters can exhibit runaway behavior, where the majority of the cluster will eventually form a single black hole with a mass much greater than the original PBH mass. These mergers can dramatically alter the PBH mass distribution and leave behind merged relic black holes that evaporate after Big Bang nucleosynthesis and yield various observational signatures, excluding parameter choices previously thought to be viable.
\end{abstract}

{
\parskip=0pt
 \maketitle
}

\section{Introduction}
\label{sec:intro}

\begin{figure*}
    \centering
    \includegraphics[width=\textwidth]{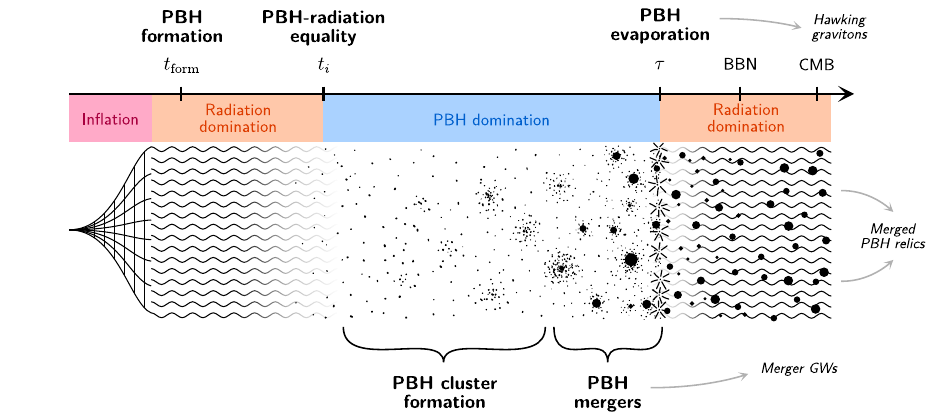}
    \caption{Example timeline describing the black hole domination scenario. At $t = t_\form$ a subdominant population of PBHs forms during radiation domination in the early universe. Since PBHs redshift like nonrelativistic matter, their energy fraction grows linearly during this era and they dominate the energy budget at $ t = t_i$, which serves as the starting point for our analysis in this paper. During PBH domination, density perturbations grow linearly with scale factor and PBH clusters begin to form. Within a range of cluster masses, PBH mergers can be fast compared to Hubble expansion and consume most of the cluster in a runaway merger. This process forms a sub-population of more massive, longer lived PBH relics which can survive past $t = \tau$ when the original PBH population evaporates due to Hawking emission. Thus, even though the original PBH population evaporates before BBN, the relic population can have important consequences for BBN and CMB observables and may also survive into the later universe. This sequence of events gives rise to gravitational waves from the PBH mergers, the Hawking emission at evaporation, and from second order scalar perturbations.}
    \label{fig:schematic}
\end{figure*}

First postulated by Hawking in 1971 \cite{HawkingGravitationallyCollapsedObjects1971}, primordial black holes (PBH) may have formed in the early universe from large density fluctuations on small scales; see Ref.~\cite{SasakiPBHsPerspectivesGW2018} for a review. If the PBH mass scale falls within $10^{17}\,\mathrm{g} < m < 10^{23}\,\mathrm{g}$, PBHs can viably accommodate the total dark matter abundance in our universe \cite{CarrConstraintsPBHs2021,GreenPBHsDMCandidate2021}; for $m >10^{23}\,\mathrm{g}$, these objects are subject to various astrophysical bounds and can account for a small fraction of dark matter. The cosmological abundance of lighter PBHs with mass $< 10^{17}\,\mathrm{g}$ is generically subject to stringent cosmological limits from the cosmic microwave background (CMB) and Big Bang nucleosynthesis (BBN)~\cite{AcharyaCMBBBNConstraints2020}.

However, these low mass limits only apply if the initial PBHs survive past BBN. If, instead, the PBH lifetime is shorter than $\sim 1\,\mathrm{s}$, they can realize a vastly wider range of initial abundances, and can even temporarily dominate the cosmic energy density. For example, if the radiation dominated early universe had a temperature $T_\form$ at the time of PBH formation, then for any initial PBH energy fraction $f_\BH$ that satisfies \cite{HooperDarkRadiationSuperheavy2019}
\begin{equation}
\label{eq:beta-dom}
f_\BH > 1.6 \times 10^{-13} \brac{10^{10} \, \mathrm{GeV}}{T_\form} \brac{10^9 \, \mathrm{g}}{m}^{3/2}\!\!,
\end{equation}
the PBH population eventually dominates the cosmic energy budget before evaporating through Hawking radiation to restore the Standard Model radiation bath. Since the lifetime of a BH is
\begin{equation}
\label{eq:tau_BH}
\tau \sim G^2 m^3 \sim 1 \, \mathrm{s} \, \brac{m}{10^9 \, \mathrm{g} }^3,
\end{equation}
the BHD scenario can only viably restore the radiation dominated universe for $m \lesssim 10^9 \, \mathrm{g}$; larger PBH mass values result in a radiation temperature below the MeV scale at the time of PBH evaporation \cite{HooperDarkRadiationSuperheavy2019}.

Once the early universe enters a regime of black hole domination (BHD), its homogeneous properties are largely set by the PBH mass and are insensitive to the details of earlier epochs. For example, assuming a monochromatic PBH mass function, there is a one-to-one relationship between $m$ and the radiation temperature resulting from evaporation \cite{HooperDarkRadiationSuperheavy2019}, the PBH merger rate \cite{SasakiPBHScenarioGW2016,RaidalGWsPBHMergers2017,ZagoracGUTScalePBHs2019,HooperHotGravitonsGWs2020}, the radiation densities of exotic new species \cite{HooperDarkRadiationSuperheavy2019,Shallue:2024hqe}, the baryon asymmetry \cite{Baumann:2007yr,MorrisonMelanopogenesisDMalmost2019,HooperGUTBaryogenesisPBHs2021,Bernal:2022pue}, and even the mass of dark matter if it is produced non-thermally through Hawking radiation \cite{LennonBHGenesisDM2018,MorrisonMelanopogenesisDMalmost2019,HooperDarkRadiationSuperheavy2019,Cheek:2021odj}.

While the existing literature on BHD has addressed many aspects of this compelling scenario, one important question remains unanswered: how does gravitational clustering affect the predictions of BHD? As in any other matter dominated phase \cite{AllahverdiFirstThreeSeconds2021}, during BHD, cosmological perturbations grow linearly with the cosmic scale factor and gravitationally bound structures begin to form. Depending on the duration of this early matter dominated era, the denser conditions in these black hole clusters could sharply enhance PBH merger rates and result in a dramatically modified mass distribution at the time of evaporation. Since clustering is purely gravitational, the details of early structure formation only depend on the time elapsed between the onset of BHD at $t = t_i$ and PBH evaporation at $t = \tau$.

In this paper, we use the Press-Schechter formalism \cite{PressFormationGalaxiesClusters1974} to study cluster formation and PBH mergers during an era of BHD. We find that, depending on the duration of BHD, PBH mergers can be very efficient in large clusters and significantly modify the initial PBH mass function. This both changes the generically UV insensitive predictions of BHD and even excludes certain regions of parameter space in which many PBHs grow through mergers to survive into the BBN and CMB eras.

Earlier work has considered clusters of cosmologically stable PBHs in the later universe after matter-radiation equality \cite{FishbachAreLIGOsBHs2017, DeLucaClusteringEvolutionPBHs2020,FrancioliniPBHMergersThree2022,DelosStructureFormationPBHs2024}, or focused on PBH mergers in the early Universe without cluster formation (the ``quasi-linear regime'') \cite{HooperHotGravitonsGWs2020,DeLucaClusteringEvolutionPBHs2020,NakamuraGWsCoalescingBH1997,IokaBHBinaryFormation1998,AliHaimoudMergerRatePBH2017,RaidalFormationEvolutionPBH2019,DeLucaHeavyPBHsStrongly2023,SasakiPBHScenarioGW2016,RaidalGWsPBHMergers2017,ZagoracGUTScalePBHs2019}, while Refs.~\cite{DeLucaHeavyPBHsStrongly2023,ChisholmClusteringPBHsII2011} explored the formation of heavier PBHs from strongly clustered lighter ones (clusteringenesis), but did not specify a mechanism for this enhanced initial clustering at PBH formation. Ref.~\cite{KimPBHReformationEarly2024} studied PBH formation from the growth of density perturbations in PBH domination but did not consider mergers in PBH clusters. Black hole mergers in clusters have also been studied in the context of building supermassive black holes in galaxies \cite{XuDynamicsMassiveBHs1994,TanakaAssemblySupermassiveBHs2009,InayoshiAssemblyFirstMassive2020}. Our analysis is complementary as we consider evaporating PBHs with $m < 10^9 \, \mathrm{g}$ and calculate their clustering behavior in the early universe in a BHD era, which has not been studied before.

This paper is organized as follows. In Section~\ref{sec:cosmo} we review the cosmological aspects of PBH domination and evaporation. In Sec.~\ref{sec:dynamics} we describe PBH cluster formation and the internal dynamics of these clusters. In Sec.~\ref{sec:runawaymergers} we calculate the evolution of PBH mass distribution due to mergers in clusters. In Sec.~\ref{sec:obs} we discuss the observational consequences of BHD with clustering, including contributions to dark matter, dark radiation, and primordial gravitational waves (GW). Throughout our analysis we use natural units where $\hbar = c = k_B = 1$.

\section{Cosmology of PBH Domination} \label{sec:cosmo}

In this section, we motivate the BHD as a generic cosmic epoch that results from enhanced small-scale curvature perturbations in the early Universe. The main results of our work do not depend on the details of how PBHs were formed or came to dominate the early universe but only depend on the initial conditions such as the initial time of BHD and the original BH mass. Our discussion follows the sequence of steps depicted schematically on the timeline in Fig. \ref{fig:schematic}, highlighting several important timescales that govern the cosmological evolution in BHD:

\begin{enumerate}
    \item \textbf{Formation:} the initial PBH population is created after inflation at some time $t = t_\form$. Although our results are presented independently of the details of the formation mechanism, if the PBHs form from collapsing inflationary perturbations during radiation domination (reviewed below in Sec. \ref{sec:form}), then at $t_\form \sim G m$ the universe contains an initial PBH energy fraction, as given in \Eq{eq:beta-dom}. We assume this population has a nearly monochromatic mass distribution.

    \item \textbf{Domination:} since the PBH energy density redshifts like non-relativistic matter, if the post-inflationary universe is radiation-dominated, the PBH energy fraction grows linearly with scale factor. Depending on the initial energy fraction and the initial PBH mass $m_i$, black hole domination may be established at some time $t_i$ before the PBHs evaporate via Hawking evaporation. Throughout our analysis, we take $t_i$ to be a free parameter.

    \item \textbf{Clustering:} since BHD is a special case of early matter domination driven by average PBH density $\bar \rho$, the cosmological density contrast $\delta = (\rho - \bar \rho)/\bar \rho$ grows linearly with the scale factor $\delta \propto a$ and gravitationally bound clusters begin to form. For a sufficiently long BHD era, a large fraction of PBHs become bound within these clusters by the time of PBH evaporation, $t = \tau$ in \Eq{eq:tau_BH}.

    \item \textbf{Merging:} as virialized clusters form during BHD, their PBH constituents begin to merge and modify the original mass distribution. If BHD lasts sufficiently long, the PBHs in large clusters merge multiple times and can grow to become orders of magnitude larger than their original size at formation.

    \item \textbf{Evaporation:} due to Hawking radiation \cite{HawkingBHExplosions1974,HawkingParticleCreationBHs1975}, black holes evaporate to create the second (final) radiation era during which BBN takes place. This occurs at a characteristic time $\tau > t_i$ (see Sec.~\ref{sec:pbhevaporation}). After evaporation, particles emitted as Hawking radiation must thermalize to a temperature $T \gtrsim $ MeV to allow for standard BBN, which requires $m \lesssim 10^9 \, \mathrm{g}$.
\end{enumerate}

For the remainder of this work, we study the predictions of BHD for an initially monochromatic PBH mass function and begin our analysis at the time of domination, neglecting the abundances of any other species at that time. Thus, throughout the paper, our free parameters will be $t_i$ and $m_i$ which uniquely determine the duration of BHD and the final radiation temperature of the universe after evaporation.

\subsection{PBH Basics}
A black hole of mass $m$ has a Hawking temperature~\cite{HawkingParticleCreationBHs1975}
\begin{equation}
\label{eq:TBH}
T_H = \frac{1}{8\pi G m} \approx 10^4 \, \mathrm{GeV} \brac{10^9 \, \mathrm{g}}{m},
\end{equation}
which governs its mass loss due to Hawking radiation
\begin{equation}
\label{eq:evaporationrate}
\frac{dm}{dt} = -\frac{\kappa \,g_{\star, H} (T_H)}{30720 \pi \, G^2 m^2} ,
\end{equation}
where $G$ is the Newton constant, $\kappa \approx 3.8$ is a graybody normalization factor, and $g_{\star, H}(T_H)$ counts all particle species with masses below $T_H$ following the prescription in Ref. \cite{HooperDarkRadiationSuperheavy2019} based on the results of Refs.~\cite{MacGibbonQuarkGluonJet1990,MacGibbonQuarkGluonJet1991}. Integrating \Eq{eq:evaporationrate} to obtain the PBH lifetime, we find
\begin{equation} \label{eq:tevap}
    \tau = \frac{10240 \pi \, G^2 m^3}{ {\kappa \, g_{\star, H}}} \approx 0.4 \, \mathrm{s} \, \brac{108}{g_{\star,H}} \brac{m}{10^9 \, \mathrm{g}}^3,
\end{equation}
where $g_{\star, H} \approx 108$ is for Standard Model particle emission at BH temperatures well above the electroweak scale. In our mass range of interest ($m < 10^9\,\mathrm{g}$), the Hawking temperature in \eqref{eq:TBH} is always in this regime, so for a PBH that forms at $t = t_\form$, the mass evolves as
\begin{equation} \label{eq:m_evap_evolution}
    m(t) = m_i \left(1 - \frac{t - t_\form}{\tau}\right)^{1/3},
\end{equation}
where $m_i$ is the initial mass at formation. Since BHD requires $t_\form \ll \tau$, individual BH evolution is insensitive to $t_\form$.

\subsection{PBH Formation}
\label{sec:form}

There are many proposed mechanisms for creating PBHs in the early universe. In addition to the gravitational collapse of horizon-scale overdensities, primordial black holes could arise from sub-horizon processes including bubble collisions during first order phase transitions \cite{LiuPBHProductionDuring2022}, the collapse of large domain walls during second order phase transitions \cite{RubinPBHsNonequilibriumSecond2000, DunskyPBHsAxionDomain2024}, scalar condensates that form $Q$-balls \cite{CotnerPBHsSupersymmetryEarly2017, Lu:2024xnb}, the collapse of cosmic strings \cite{KibbleTopologyCosmicDomains1976,VilenkinCosmicStringsPBHs2018} and dissipative dark forces \cite{FloresStructureFormationAfter2023,BramanteDissipativeDarkCosmology2024,RalegankarGravothermalizingPBHsBoson2024}.

A common feature of all sub-horizon mechanisms is that the characteristic PBH mass is governed by additional model-dependent parameters (e.g. the bubble nucleation rate during a phase transition), so generically there is no connection between the PBH mass and its time of formation --- see Ref. \cite{GreenPBHsDMCandidate2021} for a discussion of the common mechanism of PBH formation from very large overdensities entering the horizon, and how this relates to the parameters, initial PBH mass, $m_i$ and initial time of BHD, $t_i$. Note that our analysis below is insensitive to the specific PBH formation mechanism as long as the PBH lifetime is long compared to the Hubble time at formation and our results are uniquely specified by the input parameters $m_i$ and $t_i$.

\subsection{PBH Domination} \label{sec:pbhdom}

Formally, we define $t_i$ as the time when subhorizon PBH density fluctuations begin to grow, and $a_i = a(t_i)$ as the corresponding scale factor. Depending on the cosmological history, this parameter can typically be interpreted as the time when PBHs effectively dominate the total energy density of the universe. In matter domination the scale factor and Hubble rate are
\begin{equation}
    a(t) = \brac{t}{t_i}^{2/3} \! a_i
        \quad \text{and} \quad
    H(t) = \sqrt{ \frac{8\pi G \bar \rho_\BH}{3} } = \frac{2}{3t} \, ,
\end{equation}
and until $\tau$ when the PBHs evaporate, their background density can be written
\begin{equation} \label{eq:rhoBH}
    \bar \rho_\BH(a) = \frac{1}{6 \pi G t_i^2} \brac{a_i}{a}^{3} = \frac{1}{6 \pi G t^2}.
\end{equation}

The condition of PBH domination sets two fundamental constraints on the initial PBH mass and initial time. First, the PBHs must not evaporate before BHD begins at time $t_i$, so from Eq.~(\ref{eq:tevap}), the initial PBH mass must satisfy
\begin{equation}
    m_i > \brac{\kappa \, g_{\star,H} \, t_i}{10240 \pi G^2}^{1/3} \approx 10^9 \, \mathrm{g} \, \brac{t_i}{0.4 \, \mathrm{s}}^{1/3},
\end{equation}
which ensures that the evaporation rate is slow compared to the Hubble rate at the onset of BHD. Second, in order for PBHs to viably exist at the beginning of BHD, the initial PBH mass must not exceed the horizon mass\footnote{This definition follows from setting $R_M = H^{-1}$, where $R_M$ is the length scale from \Eq{eq:R_M} associated with a Fourier space top hat window function enclosing mass $M$, and solving for $M$.}
\begin{equation}
\label{eq:MH}
    M_{H} \equiv \frac{27 \pi t}{8G},
\end{equation}
 so causality requires
\begin{equation}
    m_i < M_H(t_i) \approx 10^9 \, \mathrm{g} \, \brac{t_i}{2\times 10^{-31} \, \mathrm{s}},
\end{equation}
which governs the maximum PBH mass that can form before $t_i$.

\subsection{PBH Evaporation} \label{sec:pbhevaporation}

The SM radiation density ${\bar \rho}_R$ accumulates according to the Boltzmann equations
\begin{align}
    \dot {\bar \rho}_\BH + 3 H \bar \rho_\BH &= - \, \dot{m} \, \bar n_\BH, \\
    \dot {\bar \rho}_R + 4 H {\bar \rho}_R &= \dot{m} \, \bar n_\BH,
\end{align}
where $\bar n_\BH = \bar \rho_\BH/m$ is the background PBH number density and $\dot m$ is given by Eq. (\ref{eq:evaporationrate}). In the instantaneous evaporation approximation, $\bar \rho_\BH(\tau) \approx {\bar \rho}_R(\tau)$, so using Eq.~(\ref{eq:rhoBH}) we have
\begin{equation}
    \bar \rho_\BH(\tau) = \frac{1}{6\pi G \tau^2} \approx {\bar \rho}_R(\tau) = \frac{\pi^2}{30} g_\star(T_R) \, T_R^4 \, ,
\end{equation}
where $g_\star$ is the effective number of relativistic species in the radiation bath. Since $\tau$ from \Eq{eq:tevap} is only a function of $m$, there is a one-to-one relationship between the PBH mass and the reheated radiation temperature upon evaporation \cite{HooperDarkRadiationSuperheavy2019}
\begin{equation}
    T_\mathrm{RH} = \brac{30 \bar\rho_\BH(\tau)}{\pi^2 g_\star}^{\! 1/4}
            \! \approx 1.6\,\mathrm{MeV} \brac{10^9\,\mathrm{g}}{m}^{\! 3/2},
\end{equation}
where $g_\star$ only includes Standard Model particle species. Thus, to ensure that the radiation era that follows BHD is sufficiently hot to form the observed light element abundances during BBN, $m \gtrsim 10^9\, \mathrm{g}$ is required \cite{ParticleDataGroupReviewParticlePhysics2022}.

\section{Dynamics of PBH Clusters}\label{sec:dynamics}

Once PBHs dominate the early universe, density fluctuations begin to grow and the PBHs form self-gravitating clusters. In this section, we describe this process and the internal dynamics of these PBH clusters, including nearby encounters, mergers, and other $N$-body effects.

\subsection{PBH Cluster Formation} \label{sec:cluster_formation}

If no intrinsic correlations are present between PBH positions, the initial fluctuations are described with Poissonian shot noise (for a derivation, see Appendix \ref{app:shot_noise}), with an initial power spectrum
\begin{equation}
    P_\BH(k, a_i) = \frac{1}{\bar n_\BH(a_i)} \, ,
\end{equation}
where $\bar{n}_\BH(a) = \rho_\BH(a) / m_i$ is the average number density of the PBH population. The clustering of PBHs has been extensively studied both analytically and numerically for astrophysical PBH masses in the late universe \cite{DesjacquesSpatialClusteringPBHs2018,InmanEarlyStructureFormation2019,DomenechExploringEvaporatingPBHs2021,AuclairSmallScaleClustering2024}. Here, we apply these results to the early PBH dominated era and study the cluster evolution. In principle, inflation could have induced additional fluctuation on these length scales, but as a conservative estimate of the level of PBH clustering, we only consider the power due to the shot noise. This contribution arises from the discrete nature of the PBH distribution and is an irreducible source of fluctuations on all scales larger than some ``exclusion scale,'' which accounts for the fact that PBHs cannot physically form arbitrarily close to each other \cite{DesjacquesSpatialClusteringPBHs2018}.

When a PBH density fluctuation mode with comoving wavenumber $k$ enters the horizon in PBH domination at scale factor
\begin{equation}\label{eq:a_cross}
    a_k = \brac{2}{3 k t_i}^2 a_i \, ,
\end{equation}
it begins to grow linearly with the scale factor, $\delta \propto a$ \cite{DodelsonModernCosmology}. We assume that any modes that are already inside the horizon before time $t_i$ do not grow until after $t_i$. The power spectrum for subhorizon modes after $t_i$ then takes the form
\begin{equation}
    P_\BH(k, a) = \frac{1}{\bar{n}_\BH(a_i)} \brac{a}{a_i}^2
    \begin{cases}
        1 & k > k_i\\
        (k/k_i)^4 & k < k_i
    \end{cases},
\end{equation}
where $k_i = 2 / (3 t_i)$ is the initial comoving wavenumber of the horizon scale at time $t_i$. Note that the $k$ scaling for $k < k_i$ ensures that the corresponding modes only begin to evolve after horizon crossing at $a = a_k$ in \Eq{eq:a_cross}. Just as in standard cosmology during matter domination, large-scale modes are suppressed by entering the horizon later and growing for a shorter duration.

As BHD progresses, the initial shot noise seeds the formation of PBH clusters. The clusters will grow over time as they accrete PBHs and merge with each other in a hierarchical manner. We use the Press-Schechter formalism \cite{PressFormationGalaxiesClusters1974} to compute the mass distribution of these PBH structures at a given time during BHD.

In the spherical collapse model, overdense regions will decouple from the Hubble flow and collapse to virialized objects when the local overdensity $\delta = (\rho - \bar{\rho})/\bar{\rho}$ is larger than a critical value $\delta_c$ (for a flat matter-dominated Universe, $\delta_c\approx 1.686$). To estimate how frequently such regions exceed the critical overdensity $\delta_c$, the overdensity fluctuations are usually assumed to be Gaussian-distributed with a variance $\sigma_M^2$, which is related to the matter power spectrum. More precisely, $\sigma_M^2(a)$ is the variance of the density fluctuations smoothed over a characteristic length scale of
\begin{equation}
\label{eq:R_M}
    R_M = \brac{M}{6\pi^2 \bar{\rho}}^{1/3},
\end{equation}
which corresponds to the size of a Fourier space top hat window function containing total mass $M$ on average. It can be calculated from the PBH power spectrum smoothed with the top hat window function $W(kR) = \Theta(kR)$, such that
\begin{equation}
\label{eq:sigmaM2}
\!\! \sigma_M^2 \! = \! \int_{0}^{\infty} \!\!\!\frac{d^3k}{(2\pi)^3} P(k,a) W^2(kR_M)
                = \frac{m_i}{M} \mu(M) D(a)^2 \!,
\end{equation}
where we define the usual linear growth function in matter domination $D(a) = a/a_i$ and a ``cluster mass transfer function'' $\mu(M)$ to encapsulate the effect of mass scales entering the horizon and growing at different times:
\begin{equation} \label{eq:masstransferfunction}
    \mu(M) = \begin{cases}
        \displaystyle 1 - \frac{4}{7} \frac{M}{M_{H,i}} & M < M_{H,i} \vspace{0.3em} \\
        \displaystyle \frac{3}{7} \brac{M}{M_{H,i}}^{-4/3} & M > M_{H,i}
    \end{cases}
\end{equation}
where $M_{H,i} = M_{H}(t_i)$ is the horizon mass from \Eq{eq:MH}. Thus $\mu(M) \approx 1$ for those scales which were already inside the horizon at $t_i$, and $\mu(M) < 1$ for those which enter the horizon and begin collapsing at some later time.

The Press-Schechter ansatz \cite{PressFormationGalaxiesClusters1974} relates the fraction of mass contained in all clusters more massive than $M$ to the probability of the overdensity field (smoothed at a mass scale $M$) being larger than $\delta_c$. This connects cluster properties to the statistics of cosmological density fluctuations. The differential cluster number density as a function of the cluster mass $M$ is described by the Press-Schechter mass function
\begin{equation}
    \frac{dn_\cl}{dM} = \sqrt{\frac{2}{\pi}} \frac{\bar \rho_\BH}{M} \frac{\delta_c}{\sigma_M^2} \left| \frac{d \sigma_M}{d M} \right| \exp{ \left( -\frac{\delta_c^2}{2\sigma_M^2} \right)}.
\end{equation}
It will be convenient to label clusters by the number of PBHs they contain, $N = M / m > 1$. In particular, we will often use the initial cluster number $N_i = M_i / m_i$, because the PBH mass $m$, cluster mass $M$, and cluster number $N$ may evolve over time. Although cluster properties may change, we assume that the shot noise power spectrum is seeded upon PBH formation and does not change, depending only on $m_i$. The distribution of clusters in terms of the initial PBH number $N_i$ is
\begin{equation}
\label{eq:dndN_PS}
    \frac{dn_\cl}{dN_i} \! = \! \frac{\bar n_\BH \, \delta_c}{\sqrt{2\pi \mu_i N_i^3 } D}
    \! \left(\! 1 \!-\! \frac{d \log \mu_i}{d \log N_i} \!\right)
    \exp \! \left( \!-\frac{N_i}{2\mu_i} \frac{\delta_c^2}{D^2} \!\right)\!,
\end{equation}
where $\mu_i \equiv \mu(N_i)$ is equivalent to Eq.~(\ref{eq:masstransferfunction}) with $M = N_i m_i$ as the initial cluster mass. We also define
\begin{equation}
    N_{H,i} \equiv \frac{M_{H,i}}{m_i} = \frac{27\pi t_i}{8 G m_i},
\end{equation}
as the cluster size corresponding to the horizon mass at time $t_i$, which separates different $N_i$ scaling behavior in the expression for $\mu_i$.

The Press-Schechter formalism finds that clusters of mass $M$ tend to form when the variance of PBH overdensities (smoothed on the mass scale $M$) reaches $\sigma_M(a) \sim \delta_c$. Using Eq.~(\ref{eq:sigmaM2}), clusters with initial PBH occupation number $N_i$ form at
\begin{equation} \label{eq:a_cl}
    a_\cl = \delta_c \sqrt{\frac{N_i}{\mu_i}} \, a_i
        ~~~~ \text{or} ~~~~
    t_\cl = \delta_c^{3/2} \brac{N_i}{\mu_i}^{3/4} t_i~,
\end{equation}
which are respectively the characteristic scale factor and time of cluster formation. The Press-Schechter cluster mass function encodes this behavior, particularly in the exponential cutoff scale, with smaller clusters growing first, and larger ones appearing later. Once clusters of a given mass are no longer exponentially suppressed, their comoving number density decreases with time as they are subsumed into larger clusters. Thus the era of PBH cluster formation begins with the smallest clusters ($N_i \sim 1$) forming once the scale factor grows to $a \approx 1.686 \, a_i$.

After an overdensity collapses and virializes to form a PBH cluster, its average density can be related to the universal average density at the time that it forms \cite{CoorayHaloModelsLarge2002} (the stable clustering hypothesis):
\begin{equation}
\label{eq:rhobar_cl}
    \rho_{\cl,i}
    = \Delta \, \bar \rho_\BH(a_\cl)
    = \frac{\Delta}{6 \pi G t_i^2 \delta_c^3} \brac{\mu_i}{N_i}^{3/2},
\end{equation}
where $\Delta = 18 \pi^2$ is the characteristic overdensity of a collapsed object \cite{MoBoschWhite}. For simplicity, we assume a spherical geometry with a uniform density profile. This is a conservative choice as the merger rate at the center of clusters should be much higher due to the large density, though the specific details depend on the choice of density profile. The effective radius of the PBH cluster is initially
\begin{equation} \label{eq:R_cl}
\begin{split}
    R_i
    &= \brac{3M_i}{4\pi \rho_{\cl,i}}^{1/3} \\
    &\approx 10^{6} \, \frac{r_s^\cl}{N_i^{1/6} \mu_i^{1/2}} \, \brac{m_i}{10^9\,\mathrm{g}}^{\!-2/3} \brac{t_i}{10^{-20}\,\mathrm{s}}^{\!2/3} \! ,
\end{split}
\end{equation}
where $r_s^\cl = 2 G M_i$ is the cluster's Schwarzschild radius, and the initial velocity dispersion of the cluster is
\begin{equation} \label{eq:sigma_v_cl}
\begin{split}
    \sigma_{v,i}
    &= \sqrt{\frac{GM_i}{R_i}} \\
    &\approx 10^{-3} \, N_i^{1/12} \mu_i^{1/4} \, \brac{m_i}{10^9\,\mathrm{g}}^{\!1/3} \brac{t_i}{10^{-20}\,\mathrm{s}}^{\!-1/3} \! ,
\end{split}
\end{equation}
where we adopt a Maxwellian velocity distribution with average relative velocity $v = (4/\sqrt{\pi}) \, \sigma_v$ \cite{BinneyTremaine}. Note that Eqs.~\eqref{eq:R_cl} and \eqref{eq:sigma_v_cl} are only valid if $\sigma_{v,i} \ll 1$ and $R_i \gg r_s^\cl$, which is the case for input parameters where $m_i \ll M_{H,i}$ (above the lower gray shaded region in Fig. \ref{fig:param_constraints}).

The formation and evolution of PBH clusters from shot noise has been studied with $N$-body simulations in a matter-dominated Universe \cite{InmanEarlyStructureFormation2019}, and $N$-body simulations with similar initial conditions have been performed \cite{BlancoAnnihilationSignaturesHidden2019,XiaoSimulationsAxionMinihalos2021}, which suggest that a semi-analytic formalism such as Press-Schechter can describe the evolution of PBH clusters well; see Ref.~\cite{DeLucaClusteringEvolutionPBHs2020} for a similar discussion on the semi-analytic treatment of PBH clusters.

\subsection{PBH Binary Capture}

In this section, we describe the PBH binary capture and merger rates within clusters formed during BHD. In order to form a gravitationally bound binary black hole pair, energy must be lost in an encounter between two or more PBHs. This can occur through two different channels: with only two PBHs, a very close approach can induce a large enough emission of gravitational waves to leave behind a bound system, or with three PBHs, the third body can catalyze the interaction, carrying away enough energy that it leaves the other two bound.

The relative importance of the 2-body and 3-body binary capture channels depends on the black hole mass as well as internal cluster properties such as the average BH separation and relative velocity. Different channels also tend to form binaries with very different properties. The 2-body channel favors the formation of very close binaries with high eccentricity \cite{CholisOrbitalEccentricitiesPBH2016}, close enough that they almost immediately merge on a relatively short timescale. The 3-body channel usually forms more loosely bound binaries that may be vulnerable to further disruption before they can actually inspiral and merge \cite{LightmanDynamicalEvolutionGlobular1978}.

\begin{figure}
    \centering
    \includegraphics[width=\linewidth]{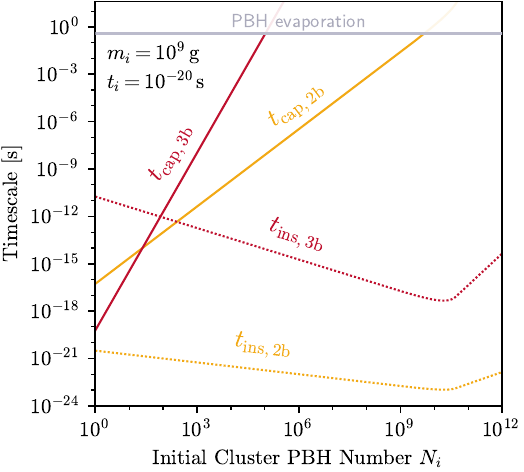}
    \caption{A comparison of the 2-body capture timescale from Eq.~\eqref{eq:t_cap_2b} and the average 2-body inspiral timescale from Eq.~\eqref{eq:t_ins_2b_avg} with the 3-body capture and inspiral timescales from Eqs.~(6) and (12) of Ref.~\cite{FrancioliniPBHMergersThree2022} for one choice of our free parameters $m_i$ and $t_i$, showing that Eq.~\eqref{eq:2b_faster_than_3b} is satisfied for all $N_i$.}
    \label{fig:2b3b}
\end{figure}

To compare the 3-body rate to the 2-body rate, consider that the rates are typically limited by either capture or inspiral, whichever is slower. By evaluating the 3-body capture and inspiral rates in Eqs.~(6) and (12) of Ref.~\cite{FrancioliniPBHMergersThree2022} and comparing to the 2-body analysis below, it can be shown that throughout the parameter space of interest, the timescales satisfy
\begin{equation} \label{eq:2b_faster_than_3b}
\max(t_\mathrm{cap,2b}, t_\mathrm{ins,2b}) \ll \max(t_\mathrm{cap,3b}, t_\mathrm{ins,3b})~,
\end{equation}
where $t_\mathrm{cap, ins}$ are respectively the capture and inspiral timescales and the 2b/3b labels refer to two and three body processes. Thus, on average, 2-body interactions lead to a quicker capture, inspiral, and merger than 3-body interactions. As a concrete example, we show in Fig.~\ref{fig:2b3b} the rates for 2-body and 3-body capture and inspiral for one choice of free parameters. For the remainder of this work we neglect 3-body interactions, which will not contribute significantly to the PBH merger rate during BHD, and instead focus solely on the 2-body interaction channel.

As two unbound black holes pass each other on a hyperbolic trajectory with impact parameter $b$ and initial relative velocity $v$ in the center-of-mass frame, they will emit ``gravitational bremsstrahlung'' radiation. Using the standard quadrupole formula for the leading order non-relativistic contribution \cite{MisnerThorneWheeler} and neglecting the backreaction on the trajectory, the total energy emitted in gravitational radiation during the encounter can be calculated, as in \cite{TurnerGravitationalRadiationPoint1977}. It is sufficient to take the parabolic limit as we expect the velocity at closest approach to be much larger than the initial relative velocity \cite{QuinlanDynamicalEvolutionDense1989,LeeNBodyEvolution1993,OLearyGWsScatteringStellar2009}, which gives
\begin{equation} \label{eq:DeltaE}
    \Delta E \approx \frac{2720 \,\pi G^7 m^8}{3\, b^7 v^7 } ,
\end{equation}
so if $\Delta E$ exceeds the total initial energy $E_i = m v^2 / 4$, then the black holes will become gravitationally bound. The cross section for forming a binary system through this process can be written
\begin{equation} \label{eq:sigma-2b}
    \sigma_\mathrm{2b} = \pi b_\mathrm{max}^2 = \brac{85\pi}{3}^{\!2/7} \frac{\pi r^2_s}{v^{18/7}},
\end{equation}
where $b_\mathrm{max}$ is the largest impact parameter for which the binding criterion $\Delta E > E_i$ is satisfied and $r_s = 2 Gm$ is the Schwarzschild radius.

The typical timescale for a single PBH in a cluster to experience binary capture via 2-body interactions is then
\begin{equation} \label{eq:t_cap_2b}
    t_\mathrm{cap,2b} = \frac{1}{n_\BH \sigma_\mathrm{2b} v},
\end{equation}
where $n_\BH = \rho_\cl / m$ is the uniform PBH number density\footnote{Note that $n_\BH$ here is not the same as $\bar n_\BH$, which represents the background average PBH number density.} in the cluster and $v = (4/\sqrt{\pi}) \, \sigma_v$ is the typical relative PBH velocity.

\subsection{PBH Mergers}

To obtain the merger rate, we first need to understand how long it takes for the PBH binaries that are formed through this process to inspiral and merge. From standard orbital mechanics, the bound black holes follow an elliptical trajectory characterized by
\begin{equation}
\label{eq:orbital}
    a_0 = -\frac{G m^2}{2 E_f}
        \quad \text{and} \quad
    e_0 = \sqrt{1 + \frac{4 E_f L_f^2}{G^2 m^5}},
\end{equation}
where $a_0$ is the semi-major axis, $e_0$ is the ellipticity, $E_f = E_i - \Delta E$ is the final energy, and $L_f$ is the final angular momentum.

The rate of inspiral over many orbital periods can be calculated based on emission of gravitational waves, similar to Eq.~(\ref{eq:DeltaE}). For a given $a_0$ and $e_0 \approx 1$, the total inspiral time from capture to merger is \cite{PetersGravitationalRadiationMotion1964,OLearyGWsScatteringStellar2009}
\begin{equation}
\label{eq:t_ins_2b}
    t_\mathrm{ins,2b} = \frac{3\, a_0^4}{170 \, G^3 m^3} \left( 1 - e_0^2 \right)^{7/2},
\end{equation}
The change in angular momentum is typically negligible \cite{OLearyGWsScatteringStellar2009}, so we approximate $L_f \approx L_i = \frac{1}{2} m b v$. For $b < b_\mathrm{max}$, the average inspiral time is
\begin{equation} \label{eq:t_ins_2b_avg}
    t_\mathrm{ins,2b}^\mathrm{avg} \approx 10.2 \, \frac{G m}{v^3},
\end{equation}
where we have used \Eq{eq:orbital} and integrated \Eq{eq:t_ins_2b} over $b \in [0, b_\mathrm{max}]$ with the appropriate weight to find the average.

To understand whether a PBH binary could be disrupted before it merges, we can consider the ratio of the typical inspiral timescale to the capture timescale in the PBH cluster environment:
\begin{equation}
\frac{t_\mathrm{ins,2b}^\mathrm{avg}}{t_\mathrm{cap,2b}} \approx 10^{-4} \, \brac{\mu_i^{5/14}}{N_i^{79/42}} \left( \frac{m_i}{10^9\,\mathrm{g}} \frac{10^{-20}\,\mathrm{s}}{t_i} \right)^{\!10/21} \!\! ,
\end{equation}
which indicates that in the parameter space of interest, particularly for large $N_i$ clusters, $t_\mathrm{ins,2b} \ll t_\mathrm{cap,2b}$, and a captured binary may be assumed to merge immediately, before any potential disruption from a third body. Thus, in our treatment, the merger rate in a PBH cluster is approximately the binary capture rate, which can be written~\cite{BirdDidLIGODetect2016,DeLucaClusteringEvolutionPBHs2020}
\begin{equation} \label{eq:Gamma_mrg}
    \Gamma_\mrg = \frac{1}{2} N n_\BH \sigma_\mathrm{2b} v,
\end{equation}
where $n_\BH$ is the average PBH number density in the cluster and $v = (4/\sqrt{\pi}) \, \sigma_v$ is the typical relative PBH velocity. The initial merger rate satisfies $\Gamma_\mrg \propto N_i^{-0.63}$ and $\Gamma_\mrg \propto N_i^{-2.1}$ for $N_i < N_{H,i}$ and $N_i > N_{H,i}$, respectively, where $N_{H,i}$ is the cluster size corresponding to the initial horizon mass. Therefore, smaller PBH clusters always have a faster merger rate than larger ones.

\subsection{Cluster Evaporation}

One major difference between PBH clusters and typical dark matter halos is that there is a lower limit on a PBH cluster mass, since $M$ cannot be smaller than $m$. The smallest clusters will only contain a few objects, so they are susceptible to $N$-body effects that lead to the ejection of their members. This is usually called cluster ``evaporation,'' not to be confused with the Hawking evaporation of black holes \cite{LightmanDynamicalEvolutionGlobular1978,BinneyTremaine,AfshordiPBHsDMPower2003,FrancioliniPBHMergersThree2022,DeLucaClusteringEvolutionPBHs2020,DeLucaHeavyPBHsStrongly2023,CeloriaLectureNotesBH2018}.

This process can be understood as follows: one of the fundamental timescales over which a gravitationally bound cluster may evolve is the dynamical timescale $t_\mathrm{dyn} = R / \sigma_v$, also known as the crossing timescale, where $R$ is the cluster radius. A related property of $N$-body systems is the relaxation timescale \cite{BinneyTremaine}
\begin{equation} \label{eq:trlx}
    t_\mathrm{rlx} \approx \frac{N}{8 \log N} \, t_\mathrm{dyn} = \frac{N}{8 \log N} \frac{R}{\sigma_v},
\end{equation}
over which many weak gravitational encounters and small deflections allow the cluster velocity distribution to equilibrate. However, some fraction $\gamma$ of the objects on the high tail of the velocity distribution may exceed the escape velocity of the cluster and will ``evaporate,'' or at least move to a more distant orbit where encounters are rare. Applying the virial theorem, it can be shown that for a Maxwellian velocity distribution the fraction that exceeds the root-mean-square escape velocity $2 \sqrt{3} \sigma_v$ is $\gamma \approx 7.4 \times 10^{-3}$ \cite{BinneyTremaine}. Then, within another $t_\mathrm{rlx}$, the cluster re-virializes, the high-velocity tail is replenished, and more objects may escape. The loss rate of black holes from cluster evaporation is \cite{AmbartsumianDynamicsOpenClusters1938,SpitzerStabilityIsolatedClusters1940,ChandrasekharDynamicalFrictionIII1943,LightmanDynamicalEvolutionGlobular1978,BinneyTremaine}
\begin{equation} \label{eq:Gamma_ev}
    \Gamma_\ev = \frac{\gamma N}{t_\mathrm{rlx}},
\end{equation}
so the time at which the entire cluster evaporates can be estimated as $t_\ev \sim N / \Gamma_\ev \sim t_\mathrm{rlx} / \gamma$, but a more careful consideration of how the cluster evolves during evaporation can be found in Sec.~\ref{sec:cluster_evap_dominates}.

\section{Runaway PBH Mergers} \label{sec:runawaymergers}

We now have the ingredients to demonstrate that cluster evaporation, in conjunction with binary mergers, can lead to the collapse of PBH clusters into a single merged object. In this section we combine all of the effects introduced in Sec. \ref{sec:dynamics} to study the internal evolution of PBH clusters during BHD. To make the problem more tractable, we make two crucial simplifications:
\begin{enumerate}
    \item We assume that the dynamics within a given cluster are well-described by its average PBH mass $\mavg = M / N$, so that we do not need to track the entire PBH mass distribution. We expect that the true distribution of PBH masses in a cluster will generally be peaked at some characteristic value, which motivates using $\mavg$.

    \item We assume that each PBH cluster can be treated as independent and isolated over the timescale relevant to its internal dynamics. This treatment is justified if the runaway merger timescale is shorter than the Hubble time.
\end{enumerate}
With these assumptions, we can write differential equations describing the evolution of the properties of a PBH cluster. The change in the total number $N$ of black holes in a cluster is
\begin{equation} \label{eq:N_evolution}
    \frac{dN}{dt} = - \Gamma_\mrg - \Gamma_\ev,
\end{equation}
which depends on the binary merger and cluster evaporation rates. The total mass $M$ of the cluster changes due to mergers (radiated away as gravitational waves), cluster evaporation, and Hawking evaporation:
\begin{equation} \label{eq:M_evolution}
    \frac{dM}{dt} = -\Delta m \, \Gamma_\mrg - \mavg \, \Gamma_\ev - N \dot{m},
\end{equation}
where $\Delta m \approx 0.05 \mavg$ \cite{LIGOScientificVirgoObservationGWsBinary2016} is the energy lost to gravitational waves in the merger of two equal mass Schwarzschild black holes and $\dot{m}$ is the Hawking luminosity of a single BH from \Eq{eq:evaporationrate}.

In the following subsections, we consider the behavior of PBH clusters in the limiting regimes where either the merger rate, or the cluster evaporation rate dominates the dynamics.

\subsection{Merging Dominates} \label{sec:merging_dominates}

If the merger rate is initially much faster than the cluster evaporation rate ($\Gamma_\mrg \gg \Gamma_\ev$), we can take the $\Gamma_\ev \to 0$ limit in Eqs.~\eqref{eq:N_evolution} and \eqref{eq:M_evolution} and combine the resulting expressions to obtain
\begin{equation} \label{eq:m_avg_evolution}
    \frac{d\mavg}{dt} = \frac{\mavg^2}{M} \left(1 - \frac{\Delta m}{\mavg}\right) \Gamma_\mrg - \dot{m},
\end{equation}
where we have used $\mavg = M/N$ and applied the quotient rule. If we further neglect the Hawking evaporation and $\Delta m$ terms (to ensure that $M$ is constant) and assume that $\Gamma_\mrg$ is independent of $\mavg$, which is true for 2-body interactions, the analytical solution to \Eq{eq:m_avg_evolution} can be written
\begin{equation}
    \mavg(t) = \frac{m_i}{1 - (t - t_\cl) / t_\mrg},
\end{equation}
which describes a cluster that forms at $t = t_\cl$ given in \Eq{eq:a_cl}
with initial condition $\mavg(t_\cl) = m_i$, and experiences negligible merging until it nears the critical ``runaway'' timescale
\begin{equation} \label{eq:t_mrg}
\begin{split}
    t_\mrg = \frac{N_i}{\Gamma_{\mrg} (N_i)},
\end{split}
\end{equation}
at which point the cluster quickly all merges into a single black hole.

\subsection{Cluster Evaporation Dominates} \label{sec:cluster_evap_dominates}

In the opposite regime, where $\Gamma_\mrg \ll \Gamma_\ev$, cluster dynamics are initially governed by ``cluster evaporation'' effects in which the object ejects some fraction of its PBH constituents over the course of its evolution. As kinetic energy is removed from the inner cluster by escaping PBHs, the remaining objects become more tightly bound and collapse inward with higher density and greater velocities, exhibiting the negative heat capacity of gravitating systems. Consequently, the dynamical and relaxation timescales become shorter, and the evaporation process accelerates, culminating in a so-called ``gravothermal catastrophe'' \cite{LyndenBellGravoThermalCatastrophe1968,BalbergSelfinteractingDMHalos2002}.

However, eventually, additional microphysics becomes relevant and arrests this catastrophe. In our scenario involving PBH clusters during BHD, binary formation and mergers become increasingly efficient as the cluster contracts, transitioning into the $\Gamma_\mrg \gg \Gamma_\ev$ regime described above in Sec.~\ref{sec:cluster_evap_dominates}, and eventually the resulting inner core merges into one large black hole. Thus, the final mass of this merged object depends on the fraction of PBHs that are ejected during the period before runaway merger occurs.

In general, it is difficult to determine how many PBHs escape and how this affects the fate of a cluster without running full $N$-body simulations with mergers. However, with some simplifications, we can attempt to estimate this final mass, accounting for how the properties of the cluster change during evaporation. Since objects evaporate from the cluster after a series of weak gravitational encounters nudge their velocity over the threshold for escape, we can assume that when they escape, they are only marginally unbound and they leave with approximately zero total energy. Thus, while the total mass of the cluster decreases, its total energy $E \sim G M^2 / R$ remains the same. With mass loss but no energy outflow, the characteristic cluster radius evolves as
\begin{equation}
    R \approx R_i \brac{M}{M_i}^2,
\end{equation}
where we have assumed that the spatial PBH distribution remains self-similar throughout this evolution. Using this radius scaling, the average cluster density then evolves as
\begin{equation}
    \rho_\cl \propto M / R^3 \propto M^{-5},
\end{equation}
and we see that, as objects evaporate, the cluster shrinks in size and becomes significantly more dense, signifying a collapse.

We can study the endpoint of this collapse behavior by considering just the cluster evaporation term in \Eq{eq:N_evolution}. Without mergers or Hawking evaporation, $\mavg$ is constant, so $R \propto N^2$. Given this scaling and the factors in $\Gamma_\ev$, the evolution of the number of PBHs can be expressed in terms of initial quantities and $N$:
\begin{equation}
    \frac{dN}{dt} = - \frac{8 \gamma \,\sigma_{v,i}}{R_i} \brac{N}{N_i}^{\!\!-5/2} \!\log N .
\end{equation}
Neglecting the logarithmic $N$ dependence, there is an analytic solution to this differential equation,
\begin{equation}
\label{eq:N_ce}
    N(t) \approx N_i \left( 1 - \frac{t- t_\cl}{t_\ev} \right)^{2/7},
\end{equation}
where $t_\cl$ from \Eq{eq:a_cl} is the characteristic time at which clusters with $N_i$ PBHs form, though generically $t_\cl \ll t_\ev$, so the dynamics are independent of this initial value. This expression describes a cluster losing objects until it reaches $N = 0$ at
\begin{equation} \label{eq:t_ev}
    t_\ev = \frac{2 N_i}{7\, \Gamma_{\ev}(N_i)},
\end{equation}
which sets the finite cluster evaporation timescale.

\subsection{Combined Treatment}
We now consider the interplay of both cluster evaporation and PBH mergers to understand the ultimate fate of a given cluster in BHD. At the time of its formation, if a cluster starts off in the merger dominated regime of Sec.~\ref{sec:merging_dominates}, it will undergo a runaway process that yields a large black hole with approximately the total mass of the initial cluster. However, if the cluster starts in the cluster evaporation regime of Sec.~\ref{sec:cluster_evap_dominates}, we must first track its evolution as it ejects objects and eventually transitions into the merger dominated regime. Our goal in this subsection is to determine the final ``relic'' mass, $m_\relic$, in a cluster that exhibits both of the regimes described above at different phases of its evolution.

Combining Eqs.~(\ref{eq:N_evolution}) and (\ref{eq:M_evolution}) and neglecting the Hawking evaporation term, we can write
\begin{equation}
\label{eq:dMdN}
    \frac{dM}{dN} = \frac{M}{N} \, \frac{\Gamma_\ev + \frac{\Delta m}{\mavg} \, \Gamma_\mrg}{\Gamma_\ev + \Gamma_\mrg},
\end{equation}
which can be solved numerically to find how the total cluster mass evolves as a function of $N$. However, an analytic solution to \Eq{eq:dMdN} is possible if we assume that the PBH mass $\mavg$ does not change significantly while cluster evaporation dominates, so that the terms in Eq.~(\ref{eq:dMdN}) depend only on a changing $N$. The merger rate defined in Eq.~(\ref{eq:Gamma_mrg}) and the cluster evaporation rate from Eq.~(\ref{eq:Gamma_ev}) evolve as
\begin{align}
    \Gamma_\mrg &= \Gamma_{\mrg,i} \, (N_i / N)^{45/14} \label{eq:gamma_mrg_scaling}\\
    \Gamma_\ev &= \Gamma_{\ev,i} \, (N_i / N)^{5/2}, \label{eq:gamma_ce_scaling}
\end{align}
where we have parametrized the rates in terms of their initial values ($\Gamma_{\mrg,i}$ and $\Gamma_{\ev,i}$ evaluated at $N = N_i$), and neglected the logarithmic $N$ dependence in $\Gamma_\ev$ inherited from \Eq{eq:trlx}. For the remainder of this work, we assume that these conditions approximate the true behavior of the system, though a proper $N$-body simulation is ultimately required for a full validation.

For most parameter choices of interest, cluster evaporation initially dominates ($\Gamma_{\ev,i} \gg \Gamma_{\mrg,i}$), but the steeper $N$ dependence of $\Gamma_\mrg$ eventually leads to a crossover at some point during cluster evaporation and contraction. After this point, the merger rate will quickly outrun any further mass loss from cluster evaporation and the system ultimately collapses into a single black hole with mass
\begin{equation}
\label{eq:m_relic}
    m_\relic = m_i \left( \frac{1 + \cfrac{\Gamma_{\mrg,i}}{\Gamma_{\ev,i}} \, N_i^{\tfrac{5}{7}}}{1 + \cfrac{\Gamma_{\mrg,i}}{\Gamma_{\ev,i}}} \right)^{\tfrac{7}{5} \left(1-\tfrac{\Delta m}{\mavg} \right)},
\end{equation}
which we obtained by inserting Eqs.~(\ref{eq:gamma_mrg_scaling}) and (\ref{eq:gamma_ce_scaling}) into \Eq{eq:dMdN}, solving the differential equation for $M(N)$, and finding the mass at which $N = 1$. As expected, a larger initial hierarchy between cluster evaporation and merger rates results in more PBH loss and a smaller final mass.

Since both initially merger-dominated and cluster evaporation-dominated clusters can experience runaway merging, but on different timescales, for general initial conditions we express the timescale of cluster collapse as
\begin{equation}
    t_\col = \min \left( t_\mrg,t_\ev \right),
\end{equation}
though, as noted previously, for generic inputs in our parameter space, we almost always find that $t_\col \approx t_\ev$. Since BHD only lasts until $t = \tau$, when most of the PBHs evaporate due to Hawking emission, relic formation due to runaway mergers only occurs in clusters for which $t_\col < \tau$.

\subsection{Merged Relics}

We now aim understand the mass distribution and abundance of these merged relic PBHs. Combining Eqs.~\eqref{eq:t_ev}, \eqref{eq:trlx}, \eqref{eq:R_cl}, \eqref{eq:sigma_v_cl}, and \eqref{eq:Gamma_ev}, the cluster evaporation time scales as $t_\ev \propto N_i^{7/4}$, so larger PBH clusters take longer to reach the point of collapse and runaway merger. Thus, the largest clusters that undergo runaway mergers enter this regime near the very end of BHD with a collapse time $t_\col(N_i) \approx \tau$. If cluster evaporation initially dominates ($t_\col \approx t_\ev$), this corresponds to clusters with an initial number of PBHs
\begin{equation} \label{eq:N_col_max}
    N_\col^\mathrm{max} \approx 10^{10} \left\{ \begin{array}{l}
        \displaystyle 12 \, \brac{m_i}{10^9\,\mathrm{g}}^{\!12/7} \brac{t_i}{10^{-20}\,\mathrm{s}}^{\!-4/7} \vspace{0.3em}\\
        \displaystyle 6.5 \, \brac{m_i}{10^9\,\mathrm{g}}^{\!8/11}
    \end{array} \right. \!\!\! ,
\end{equation}
where the first and second cases apply for $N_\col^\mathrm{max} < N_{H,i}$ and $N_\col^\mathrm{max} > N_{H,i}$, respectively, reflecting the sub/super-horizon divide in \Eq{eq:masstransferfunction}. In terms of our free parameters, the approximate dividing line between these cases can be written $t_i \approx 5 \times 10^{-20}\,\mathrm{s} ~ (m_i / 10^9\,\mathrm{g})^{19/11}$. In these expressions, we have taken a fiducial value of $\log N_i \approx \log{10^{11}}$ and neglected the linear part of $\mu(N_i)$ in the initially sub-horizon $N_i < N_{H,i}$ regime.

At $t = \tau$, the end of BHD, all clusters for which $N_i < N_\col^\mathrm{max}$ will have merged into a single black hole with mass $m_\relic$ given in \Eq{eq:m_relic}, while all larger clusters with $N_i > N_\col^\mathrm{max}$ experience negligible mergers and their constituent PBHs evaporate. Thus, the most massive merged relic PBH will have a mass given by $m_\relic(N_\col^\mathrm{max})$ which, taking the $\Gamma_{\ev,i} \gg \Gamma_{\mrg,i}$ limit, can be expressed
\begin{equation} \label{eq:m_relic_max}
    m_\relic^\mathrm{max} \approx 10^{15}\,\mathrm{g} \left\{ \begin{array}{l}
        \displaystyle 24 \, \brac{m_i}{10^9\,\mathrm{g}}^{\!3.53} \brac{t_i}{10^{-20}\,\mathrm{s}}^{\!-1.27} \vspace{0.3em}\\
        \displaystyle 1.4 \, \brac{m_i}{10^9\,\mathrm{g}}^{\!1.35}
    \end{array} \right. \!\!\! ,
\end{equation}
for the same cases as \Eq{eq:N_col_max}. There will also be a subdominant distribution of smaller relics. Note that for the second case, the result is effectively independent of $t_i$ because these relics are formed from clusters seeded by fluctuations that entered the horizon and began growing long after the start of BHD, making them insensitive to the initial time of the era.

\begin{figure*}
    \centering
    \includegraphics[width=0.49\linewidth]{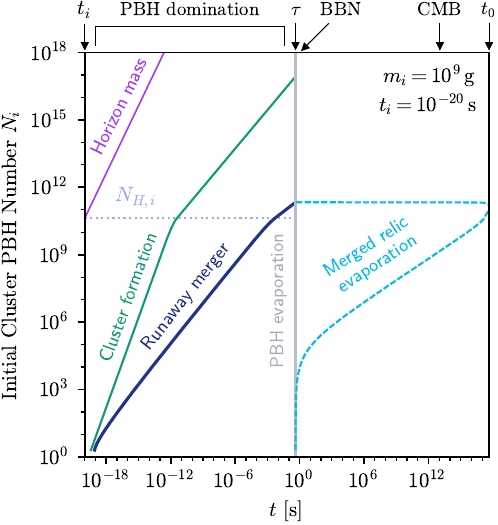}
    \hspace{0.009\linewidth}
    \includegraphics[width=0.49\linewidth]{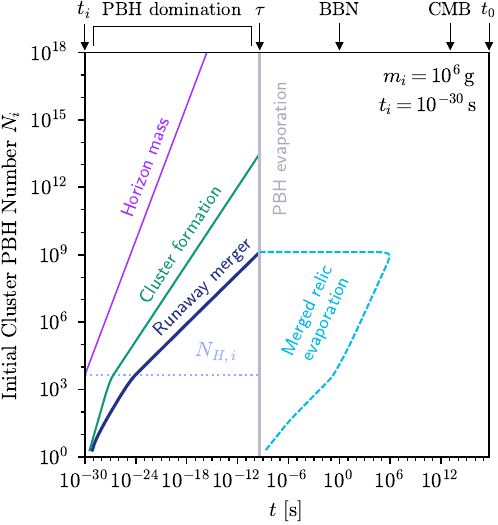}
    \caption{A summary of important timescales in our scenario for a range of initial cluster sizes $N_i$, for two choices of free parameters $m_i, t_i$. Relevant cosmic events are labeled at the top, including BHD, BBN, CMB, and the present time. The purple~\colorindicator{AA33FF} line gives the cluster size corresponding to the mass contained within the horizon. The green~\colorindicator{119977} curve shows the size of the PBH cluster which can form at a given cosmic time. At a later time, indicated by the dark blue~\colorindicator{223388} runaway merger curve, PBH clusters of a given size collapse and merge to a single BH. Finally, this merged relic evaporates at the time of the blue~\colorindicator{00BBEE} dashed curve. The light blue~\colorindicator{99AAFF} $N_{H,i}$ line divides scales that were sub- and super-horizon at $t_i$.}
    \label{fig:timescales}
\end{figure*}

Figure~\ref{fig:timescales} shows the critical times associated with various processes of PBH clustering and merging. Sliced horizontally, it can be read as a timeline for each of the cluster sizes displayed, indicating the times at which the corresponding perturbations enter the horizon, clusters form, runaway mergers begin, and when their merged relics undergo Hawking evaporation. Note that small clusters have a shorter merger timescale, mostly because they form earlier than larger PBH clusters. The clusters that form near the upper portion of the green curve (larger than $N_\col^\mathrm{max}$) do not enter the runaway merger regime before $t = \tau$, and do not leave behind PBH relics.

\begin{figure*}
    \centering
    \includegraphics[width=0.497\linewidth]{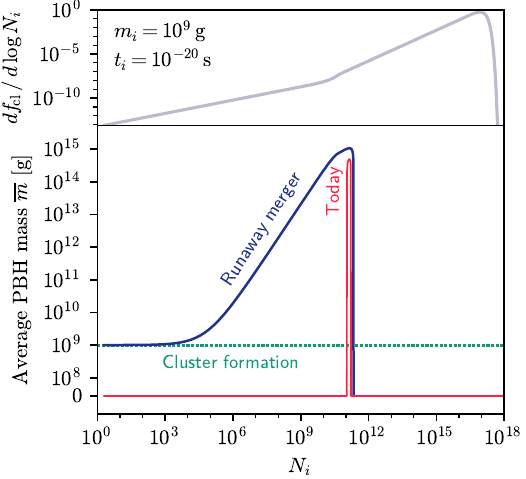}
    \includegraphics[width=0.497\linewidth]{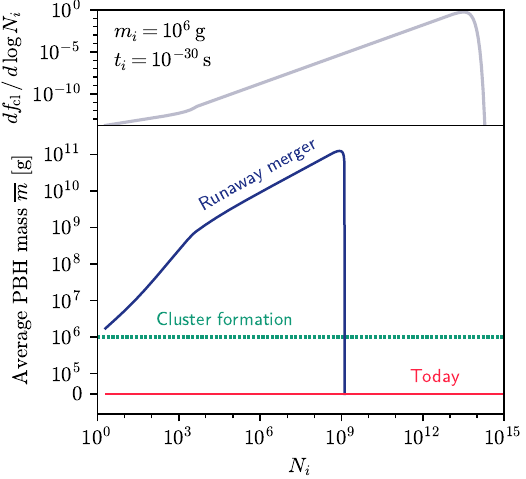}
    \caption{Average PBH mass as a function of initial cluster PBH number $N_i$ at different times, for two choices of initial parameters $m_i$ and $t_i$. Initially at cluster formation ($t=t_\cl$), the average mass is $m_i$ (green~\colorindicator{119977} line), but eventually some clusters undergo runaway mergers and their average mass grows, finally reaching the mass shown by the dark blue~\colorindicator{223388} line at $t = t_\col$.
    The red~\colorindicator{FF2244} curve shows the current remaining relic masses after the cosmologically unstable ones have evaporated away by the present day ($t=t_0$). Note that for the parameters on the right, all relics evaporate before the present day. The top panel on either side shows the fractional distribution of PBHs in each cluster size at the time of PBH evaporation $\tau$ in gray~\colorindicator{BBBBCC}.}
    \label{fig:mass_vs_Ni}
\end{figure*}

Figure~\ref{fig:mass_vs_Ni} shows the average PBH mass $\mavg = M/N$ of clusters of various initial sizes $N_i$ at cluster formation, runaway merger, main PBH evaporation, and the present day. Note that post-BHD Hawking evaporation modifies the mass distribution of the relics after the bulk of the relic population has also evaporated, leaving behind only its most massive members, as indicated by the difference between the blue and red curves on the left panel. Fig.~\ref{fig:mass_vs_Ni} also illustrates that the cluster size forming at PBH evaporation, which is also the most abundant cluster size, is significantly larger than $N_\col^\mathrm{max}$, so only a small fraction of all PBHs will merge and survive the initial evaporation. This is usually necessary to produce a cosmology consistent with our universe; the majority of the PBH population should not merge to masses above $10^9\,\mathrm{g}$ in order to reheat the Universe hot enough for BBN. Also note that the Eqs.~\eqref{eq:N_col_max} and \eqref{eq:m_relic_max} can be directly compared to the values shown in the left plot.

To calculate the merged relic PBH mass distribution at the end of BHD, we integrate over the Press-Schechter cluster number density from \Eq{eq:dndN_PS} so that
\begin{align} \label{eq:dndm_relic}
    \frac{dn_\BH}{dm}(\tau) &= \int_1^{N_\col^\mathrm{max}} \! dN_i \, \frac{dn_\cl}{dN_i}(\tau) \, \delta\big( m - m_\relic(N_i) \big) \nonumber \\
            &\quad + \int_{N_\col^\mathrm{max}}^{\infty} \! dN_i \, N_i \frac{dn_\cl}{dN_i}(\tau) \, \delta\left( m - m_i \right) \nonumber \\
        &\approx \sum_{N_i = \{N_i(m)\}} \frac{dn_\cl}{dN_i} \left| \frac{\partial m_\relic}{\partial N_i} \right|^{-1} \nonumber \\[0.3em]
            &\quad + \bar{n}_\BH(\tau) \, \delta\left( m - m_i \right),
\end{align}
where the sum is over all $N_i$ which solve the equation $m_\relic(N_i) = m$ and are less than $N_\col^\mathrm{max}$. In the second term we approximate the number density of non-merged PBHs as the total PBH number density $\bar{n}_\BH$ because the merged relics are only a small fraction of the total PBH mass. Although we assumed that all PBHs initially have the same mass $m_i$, a non-monochromatic mass distribution arises because of the distribution of cluster masses generating different merged relics. Over time, subsequent Hawking evaporation modifies this distribution, as in \Eq{eq:m_evap_evolution}, and the number density of relics dilutes as $\propto a^{-3}$.

\subsection{Important Caveats}
\label{sec:caveats}
The results presented in the previous subsections are based on the approximate solution in \Eq{eq:m_relic}, which neglects several potentially important physical effects:
\begin{enumerate}
    \item \textbf{Mass dependence of rates:} The scaling of $\Gamma_\mrg$ and $\Gamma_\ev$ in Eqs.~(\ref{eq:gamma_mrg_scaling}) and (\ref{eq:gamma_ce_scaling}) neglects the effect of changes in $M$ (or equivalently changes in $\mavg$), so the analytic solution for $m_\relic$ does not fully capture the final stages of runaway mergers. However, we have verified that this approximate solution is consistent with a numerical solution to Eq.~(\ref{eq:dMdN}) that takes this effect into account within our parameter space of interest.

    \item \textbf{Relativistic effects:} Over the course of cluster evolution, the PBH velocity dispersion and core density generically increase. Thus, our merger formalism might eventually receive important relativistic corrections, which can affect the timescale of collapse. For example, as the cluster shrinks, its characteristic radius can approach its own Schwarzschild radius, thereby forming a larger black hole without the need of additional two-body merger events. However, it is expected that these effects would only serve to accelerate cluster collapse, so our treatment here is conservative.

    \item \textbf{Gravitational bremsstrahlung:} Our treatment here conservatively neglects the additional energy dissipation from gravitational bremsstrahlung \cite{QuinlanDynamicalEvolutionDense1989}, which only serves to hasten cluster collapse; we also neglect the effects of post-merger recoils due to GW emission.

    \item \textbf{Cluster interactions:} In our analysis, we employ the Press-Schechter formalism which statistically characterizes the cluster mass function during BHD. For simplicity, we treat each cluster in isolation and ignore the possible consequences of cluster-cluster interactions, which might affect the evolution of individual objects. For example, the internal PBH cluster dynamics described in Sec. \ref{sec:dynamics} might be affected by cluster mergers or accretion, which change the properties of the host object. However, from Figure \ref{fig:timescales} (and the arguments\footnote{Although Ref. \cite{DeLucaClusteringEvolutionPBHs2020} studies PBH merging and clustering in the late universe, many of the parametric relations they derive are also applicable to our scenario due to the purely gravitational dynamics in both cases.} in Ref. \cite{DeLucaClusteringEvolutionPBHs2020}) the cluster formation rate is generically fast compared to the merger rate for any given cluster. Thus, when larger clusters form, they are rarely ``pre-enriched'' with significant numbers of merged PBHs.

    \item \textbf{Ejected merged PBHs:} We also don't account for the population of black holes that underwent some mergers but were ejected in the final stages of cluster collapse. Including this effect would enhance the mass distribution below the peak mass.

    \item \textbf{Non-uniform cluster profile:} In Eqs.~\eqref{eq:Gamma_mrg} and \eqref{eq:Gamma_ev} we assumed a uniform spherical density profile for PBHs in a cluster. Considering a realistic profile might lead to enhancement of merger rates in the denser core.

    \item \textbf{Primordial binaries:} Some binaries may form in the background before clusters can form, by virtue of their initial configuration. These have previously been explored extensively, but Ref.~\cite{DelosStructureFormationPBHs2024} finds some possible implications for gravothermal cluster evolution in a late universe scenario.

    \item \textbf{Monochromatic PBH mass function:} Throughout our analysis, we assume that the initial PBH population at $t = t_i$ is monochromatic. If PBHs form due to the collapse of primordial density fluctuations, this population corresponds to a large, highly-localized enhancement to the primordial power spectrum. In realistic inflationary models, such features can arise from  multi-field dynamics or from temporary departures away from slow-roll evolution (see Ref. \cite{Martin:2013tda} for a review). In general, these features can yield a wide range of PBH mass functions, though the details are highly model-dependent. However, if the departure from slow-roll inflation is short compared to the Hubble time during this epoch, then the power spectrum is only modified on a narrow range of scales and the PBH mass function is narrow, as in the scenario we consider.
\end{enumerate}
Since some of these effects may impact our conclusions below, it is important to follow up this analysis with a dedicated $N$-body simulation, which is beyond the scope of this paper; we leave this effort to future work.

\section{Observables and Constraints}
\label{sec:obs}

In Sec.~\ref{sec:cosmo}, we outlined three viability requirements on the parameters $m_i$ and $t_i$. In order to allow for a BHD era and recover standard cosmology after Hawking evaporation at $ t = \tau$, we demand that (1) the PBH lifetime $\tau$ exceeds the Hubble time at $t_i$, (2) the PBH mass must fit inside the horizon before $t_i$, and (3) the SM reheat temperature after PBH Hawking evaporation must exceed the $\sim$ MeV scale to allow for successful BBN. In this section, we broaden the earlier discussion to include the late-time predictions of BHD (e.g. metastable PBH relics from mergers) and the other observational bounds that constrain the input parameters $m_i$ and $t_i$.

\subsection{Merged Relics as Dark Matter Components} \label{sec:relicDM}

The BHD dynamics outlined in Sec. \ref{sec:dynamics} predict that for many choices of initial parameters ($m_i, t_i$), there is an appreciable population of merged relics with significantly larger masses than the original PBH population that drives BHD, $m_\relic \gg m_i$. Consequently, these objects are longer lived and generically survive past BBN to affect the later universe.

There are two qualitatively distinct scenarios for relic masses. For intermediate masses $10^9\,\mathrm{g} < m_\relic < 10^{15}\,\mathrm{g}$, \textit{evaporating relics} evaporate after BHD and are not constrained by late-time observations. However, because of the energy they inject while evaporating, they do face stringent limits from the successful predictions of standard BBN and CMB cosmology \cite{GreenPBHsDMCandidate2021,CarrConstraintsPBHs2021}. For $m_\relic > 10^{15}\,\mathrm{g}$, \textit{metastable relics} can be cosmologically metastable and account for an appreciable fraction of the dark matter density. This population is constrained by late-time observations including lensing, gravitational waves, and particle fluxes from Hawking evaporation, although there is an ``asteroid mass'' window between $10^{17} - 10^{23}\,\mathrm{g}$ with no current constraints \cite{GreenPBHsDMCandidate2021,CarrConstraintsPBHs2021}.

Overall, light PBHs are constrained by their Hawking evaporation with the emission of standard model particles. For the mass range of $10^{10}-10^{13}\,\mathrm{g}$, the PBH evaporation is constrained from its energy injection during the Big Bang nucleosynthesis \cite{CarrNewCosmologicalConstraints2010,CarrConstraintsPBHs2021}. If the PBH mass is slightly larger, within a mass window of $10^{13}-10^{15}\,\mathrm{g}$, the Hawking evaporation will be constrained by CMB anisotropies \cite{AcharyaCMBBBNConstraints2020}, late time extragalactic gamma-ray background (EGB) \cite{CarrNewCosmologicalConstraints2010,CarrConstraintsPBHs2021}, and cosmic rays ($e^\pm$) in Voyager \cite{Boudaud:2018hqb}.

To apply existing observational constraints to these relics, it is convenient to define a present day merged relic BH fraction
\begin{equation}
\label{eq:fBHdef}
    f_\BH \equiv \frac{\rho_\BH}{\rho_\DM},
\end{equation}
where $\rho_\DM$ is the dark matter energy density today and $\rho_\BH$ is the energy density of the BH population. Note that even though evaporating relics disappear before the present day, the quantity in \Eq{eq:fBHdef} can be defined in terms of the density that they would have today if they did not evaporate. It is also convenient to define a differential PBH fraction according to
\begin{equation}
    \frac{d f_\BH}{d \log m} = \frac{m}{\rho_\DM} \frac{d n_\BH}{d \log m},
\end{equation}
such that their mass distribution can be constrained using the criterion \cite{BellomoPBHsDMConverting2018}:
\begin{equation} \label{eq:fBHlimit}
    \int_{-\infty}^\infty d\log m \, \frac{1}{f_\BH^\mathrm{max}(m)} \,
    \frac{df_\BH}{d\log m} < 1,
\end{equation}
where $f_\BH^\mathrm{max}(m)$ is the observational limit on the PBH fraction assuming a monochromatic mass function.

\begin{figure*}
    \centering
    \includegraphics[width=0.497\linewidth]{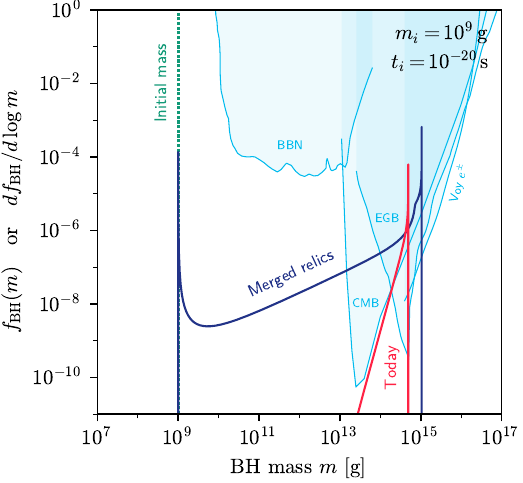}
    \includegraphics[width=0.497\linewidth]{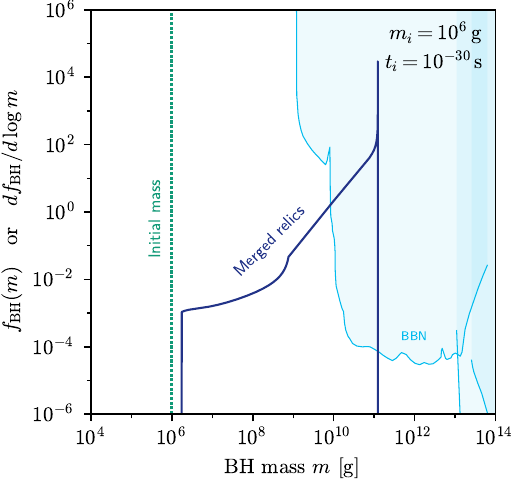}
    \caption{The mass function of merged relic PBHs for two choices of free parameters. Initially, all the PBHs have the same mass $m_i$, shown in green~\colorindicator{119977}. The blue~\colorindicator{223388} and red~\colorindicator{FF2244} curves show the initial and present-day mass functions of merged relics, which clearly shows the mass growth from runaway mergers. Superimposed in blue~\colorindicator{00BBEE} are observational bounds on evaporating BHs from BBN \cite{CarrConstraintsPBHs2021}, CMB \cite{AcharyaCMBBBNConstraints2020}, the extragalactic gamma-ray background (EGB) \cite{CarrConstraintsPBHs2021} and Voyager $e^\pm$ \cite{Boudaud:2018hqb}. Both choices of PBH parameters are ruled out by observations because they produce too many evaporating relic BHs, particularly during BBN or CMB emission. Note that the plot visually compares limits on $f_\BH$ to the logarithmic mass distribution $df_\BH / d \log m$, but the proper comparison is made with \Eq{eq:fBHlimit}.}
    \label{fig:mass_distribution}
\end{figure*}

Figure~\ref{fig:mass_distribution} presents the mass distribution of merged relic black holes at several points in time. The merged relics, which mostly have masses smaller than $10^{15}\,\mathrm{g}$, will evaporate and inject Standard Model radiation into the Universe during the BBN and CMB eras, which is strongly constrained. Note that for PBHs with masses above $10^{15}\,\mathrm{g}$, the evaporation timescale is long enough that they have not completely evaporated, but they are in the process of doing so, resulting in a new distribution of lighter ($m < 10^{15}\,\mathrm{g}$) BH relics in the current Universe, as presented in the red curve in Fig.~\ref{fig:mass_distribution}.

\begin{figure*}
    \centering
    \includegraphics[width=0.492\linewidth]{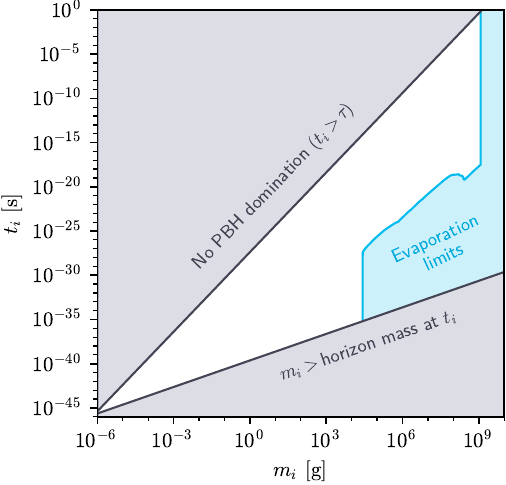}
    \hspace{0.005\linewidth}
    \includegraphics[width=0.492\linewidth]{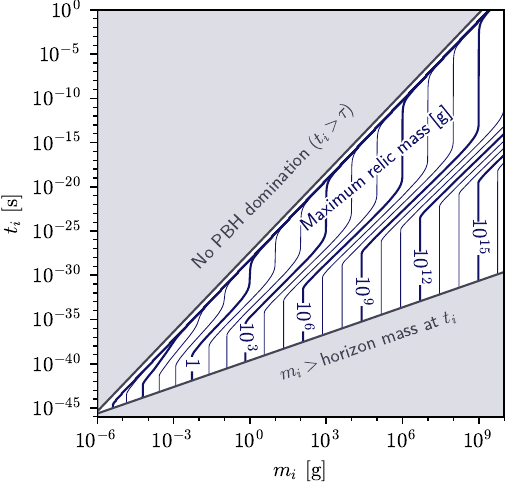}
    \\[1em]
    \includegraphics[width=0.492\linewidth]{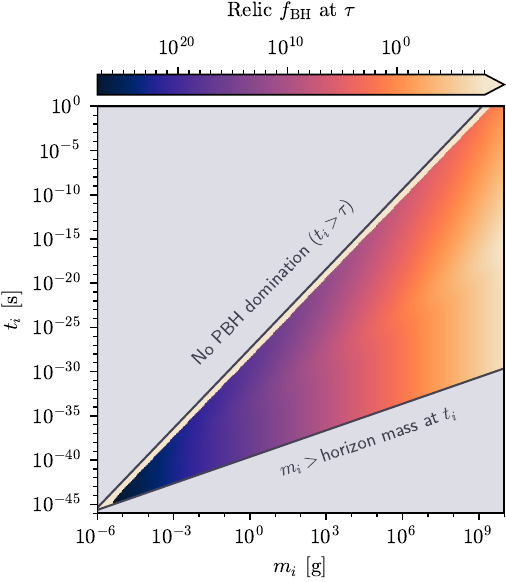}
    \hspace{0.005\linewidth}
    \includegraphics[width=0.492\linewidth]{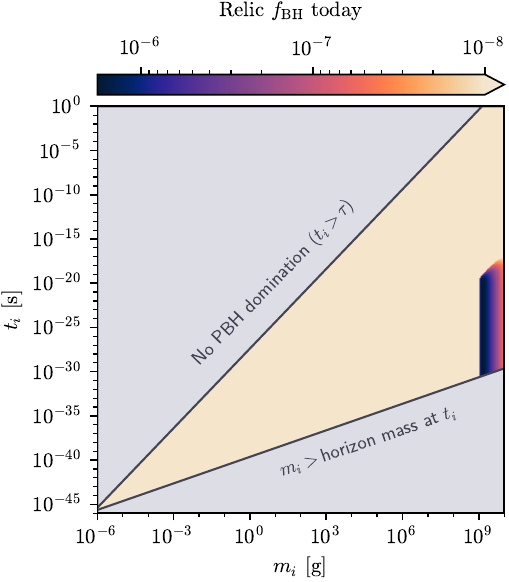}
    \caption{In this set of plots, we present the $(m_i, t_i)$ parameter space of PBH domination, and various phenomenological results. The gray regions are not physically viable as the PBHs either evaporate before initial domination or have an initial mass larger than the horizon mass. Elsewhere, the parameter space allows for BH domination and clustering. We present different information on different plots. On the upper left, we plot the parameter space ruled out by observational bounds on Hawking evaporation from the relics that form for a given $m_i$, $t_i$ combination. On the upper right, we present contours of the maximum mass of the final merged PBH relics. On the lower left, we plot the relic BH density compared to the dark matter density ($f_\BH$) at $t = \tau$, just after the original PBH population evaporates and only the relics are left. Accounting for the evaporation of BHs, on the lower right we plot $f_\BH$ of merged relics at the present time.}
    \label{fig:param_constraints}
\end{figure*}

The top left panel of Fig.~\ref{fig:param_constraints} shows our main result: the constraints on parameter space from applying the criterion in \Eq{eq:fBHlimit} to the input parameters $t_i$ and $m_i$ that characterize BHD. Note that for some free parameter values, $f_\BH \gg 1$ initially, as shown in the lower left panel of Fig.~\ref{fig:param_constraints}. However, the relic BHs in these cases are insufficiently massive to survive to the present day, so they can viably account for $f_\BH > 1$ at early times and then evaporate away well before matter-radiation equality, ensuring that dark matter is not overproduced. 

The shape of mass contours in the top right panel of Fig.~\ref{fig:param_constraints} can be understood from the different cases in \Eq{eq:m_relic_max}. The lower vertical contours correspond to parameter space where $N_\col^\mathrm{max} > N_{H,i}$, such that the largest merged relic mass is independent of $t_i$. The slanted middle region of these contours corresponds to the opposite regime, where $N_\col^\mathrm{max} < N_{H,i}$ and $m_\relic^\mathrm{max}$ depends on both $m_i$ and $t_i$. The upper vertical portion corresponds to the regime in \Eq{eq:m_relic} where $m_\relic^\mathrm{max} \approx m_i$ because cluster evaporation ejects all of the PBHs before any mergers can occur, so the mass distribution is largely unchanged by mergers.

In our scenario with a shot-noise power spectrum, it is challenging to form merged relics in the asteroid mass window. Furthermore, the region of parameter space where it is possible to produce such relics that still exist today (shown in the lower right plot of Fig.~\ref{fig:param_constraints}) is already ruled out by evaporation constraints.

\subsection{Gravitational Waves}

Another observable in this scenario is gravitational waves, which comes from several distinct sources as summarized below.

\textbf{Mergers in Clusters:} In Ref. \cite{HooperHotGravitonsGWs2020}, it was found that, under specific assumptions about PBH relative velocities and initial separations during the BHD era, GWs from the PBH mergers in the homogeneous background can account for an appreciable contribution to $\DNeff$; there was no attempt to consider clustering or merging during this matter dominated phase. However, this earlier analysis did not provide any physical rationale for why the relative velocities or separation distances should have the specific values that were chosen in that analysis. This source of theoretical uncertainty has important qualitative implications for their analysis: for example, if the PBHs are formed with negligible initial velocities, in the treatment of Ref. \cite{HooperHotGravitonsGWs2020}, their merger rates would be negligible and there would be no GWs produced from BHD beyond gravitons from Hawking radiation discussed below.

By contrast, in our treatment we only assume that the initial PBH population has a shot noise power spectrum with negligible initial velocities at the time of formation, as would be expected from the spherical collapse of perturbations. Thus, with these minimal assumptions, the only PBH mergers in our treatment occur within the clusters that form during BHD, so unlike Ref. \cite{HooperHotGravitonsGWs2020}, here only a small  fraction of PBHs participate in the runaway mergers described in Sec. \ref{sec:runawaymergers}. Since the mass fraction converted to GW per merger is approximately $\Delta m/m \sim 0.05$ \cite{LIGOScientificVirgoObservationGWsBinary2016}, the resulting contribution to dark radiation is negligible compared to the future CMB-S4 sensitivity of $\DNeff \approx 0.02$ \cite{CMBS4ScienceBook2016}. However, we note that our treatment is conservative as there will also generically be additional mergers from the unclustered background population as in Ref. \cite{HooperHotGravitonsGWs2020}, but characterizing these effects requires additional assumptions about initial velocities, which we do not make here.

Furthermore, assuming equal mass PBH mergers, the present day peak frequency of stochastic gravitational waves produced in these mergers is approximately \cite{FlanaganMeasuringGWsBinary1998,AjithInspiralMergerRingdown2011,HooperHotGravitonsGWs2020}
\begin{equation}
\label{eq:fGW}
    f^0_\mathrm{GW} \approx 10^{18}\,\mathrm{Hz} \, \brac{10^9\,\mathrm{g}}{m} \brac{\mathrm{MeV}}{T_\mathrm{RH}} \brac{g_{*,S}^\mathrm{RH}}{g_{*,S}^0 }^{1/3}\!,
\end{equation}
where $g_{*,S}$ is the number of entropic degrees of freedom, an RH label denotes a quantity at reheating ($t = \tau$), a $0$ label represents a present day quantity, and $m$ is the characteristic mass of the PBHs that undergo runaway mergers. Although the peak frequency in \Eq{eq:fGW} is far outside of any known observational technique, as noted in Ref. \cite{HooperHotGravitonsGWs2020}, the infrared tail of this stochastic background might be within the reach of the Big Bang Observer \cite{SetoPossibilityDirectMeasurement2001,YagiDetectorConfigurationDECIGOBBO2011}.

\textbf{Hawking Gravitons:} As noted in Refs. \cite{HooperDarkRadiationSuperheavy2019,HooperHotGravitonsGWs2020,CheekRedshiftEffectsParticle2022,CheekEvaporationPBHsEarly2023}, Hawking evaporation yields a population of gravitons with energy of order $T_\BH$. In the present universe, these gravitons constitute a stochastic gravitational wave background at high frequencies and also contribute to $\DNeff$. The late-time GW frequencies are generically too high for any current or future GW measurement technology, but depending on the spin and mass distribution, their contribution to the radiation density may be observable with CMB S4 \cite{CMBS4ScienceBook2016}. Since this effect depends almost entirely on Hawking evaporation from the unmerged PBH population at $t = \tau$, our treatment of clustering here does not affect these predictions except through negligible corrections.

\textbf{Secondary GWs:} Another source of GWs during BHD arises from the inhomogeneous PBH background, which is distinct from the clustering considered in this paper. At second order in perturbation theory, the scalar perturbations from the PBH shot noise power spectrum source GWs during BHD \cite{PapanikolaouGWsUniverseFilled2021,BaumannGWSpectrumInduced2007,DomenechGWConstraintsPBH2021,EggemeierStochasticGWsPostinflationary2023,EbadiGWsStochasticScalar2024}. Since the background cosmology considered here is identical to those of earlier studies, our scenario inherits this contribution to the GW density \cite{DomenechScalarInducedGWs2021,FernandezStochasticGWsEarly2024}. Furthermore, as shown in Refs \cite{DomenechGWConstraintsPBH2021,InomataEnhancementGWsInduced2019,InomataGWProductionRight2020}, the rapid process of Hawking evaporation in a PBH dominated universe also contributes to the GW density.

\section{Conclusion}

Since PBHs always redshift like non-relativistic matter, they will generically dominate the energy density of the early universe, even if their original density is negligibly small compared to that of radiation. For PBHs with masses satisfying $m_i > 10^9\,\mathrm{g}$ (with lifetimes $ \tau > 1\,\mathrm{s}$), such an epoch would spoil the successful predictions of standard BBN and is robustly excluded. However, for lighter PBHs with shorter lifetimes, Hawking radiation emitted at the PBH evaporation time thermalizes to viably restore a universe with $T > \mathrm{MeV}$, which is required for BBN. In this lower mass range, BHD is a viable attractor solution for a wide range of initial conditions and there are few known constraints on this scenario.

However, during this matter dominated era, cosmological perturbations grow linearly with scale factor and PBHs form self-gravitating virialized clusters. In this paper we have studied the fate of these clusters and found several characteristic regimes that govern their evolution. Assuming a shot noise spatial distribution of monochromatic PBHs at the time of formation, the clustering predictions of BHD are entirely characterized in terms of two free parameters: the time at which BHD begins ($t_i$) and the PBH mass scale ($m_i$). For generic choices of these inputs, the clusters that form during BHD initially evolve via ``cluster evaporation'' by ejecting PBHs through $N$-body effects. However, this behavior results in smaller, denser clusters, which are more favorable for PBH mergers. Eventually, clusters enter a ``runaway merger'' regime in which the entire object collapses into a single ``relic'' PBH of mass $m_\relic \gg m_i$.

Relics that fall within the $m_\relic \sim 10^9 - 10^{15}\,\mathrm{g}$ mass range can survive well past BHD and influence the later universe. Since these PBHs evaporate during well-studied cosmological eras, their abundance is sharply constrained by BBN, CMB, photon, and neutrino flux measurements. In this paper we have used these observables to place the first ever limits on the free parameters that define BHD $(t_i, m_i)$.

Intriguingly, relics with masses satisfying $m_\relic > 10^{15}\,\mathrm{g}$ are cosmologically metastable and can account for a fraction of the present day dark matter density. Unlike other mechanisms that form metastable PBHs (e.g. collapsing primordial density fluctuations), here the late-time abundance is \textit{not} exponentially fine-tuned. Indeed, because BHD is an attractor solution for nearly any initial PBH population after inflation, these metastable relics also arise from our mechanism for a wide range of initial values $t_i$ and $m_i$.

However, if the initial PBHs have a shot noise power spectrum, we find that relics cannot accommodate the full dark matter abundance. We find that, by assembling relic PBHs in the viable dark matter window ($m_\relic \sim 10^{17} - 10^{23}\,\mathrm{g}$) with the observed DM abundance, there is an irreducible tail of evaporating relics at lower masses ($10^9 - 10^{15}\,\mathrm{g}$) whose abundance is excluded by the BBN and CMB limits. Thus, under our minimal assumptions, this mechanism cannot account for any viable present day PBH abundance. While a different ansatz for the initial PBH power spectrum (e.g. more initial clustering than shot noise predicts) might result in enhanced merging, it remains to be seen whether that could allow for a viable dark matter formation scenario. We leave such analyses for future work.

Finally, we note that our results rely on several assumptions about PBH cluster evolution discussed in Sec. \ref{sec:caveats}. Notably, we treat all of the PBHs in a cluster as evolving collectively and simultaneously, and we assume that gravitational bremsstrahlung, cluster-cluster interactions, and relativistic effects (among others) do not significantly affect our results. While we have presented plausible arguments for why our results should not be affected in most of these cases, many of these effects are not analytically tractable and may, nonetheless, allow for different conclusions to hold; especially when multiple effects are combined. Thus, in order to robustly characterize cluster evolution and relic formation during BHD, a full $N$-body simulation will ultimately be necessary. Running such a simulation is beyond the scope of the present work, but we strongly encourage future efforts to better understand this simple, yet physically rich cosmology.

\bigskip

\section*{Acknowledgments}
We thank Nick Gnedin, Antonio Riotto,
and Lian-Tao Wang for helpful conversations.
HX and GK are supported by Fermi Research Alliance, LLC under Contract DE-AC02-07CH11359 with the U.S. Department of Energy.

\section*{Data Availability}

The code and data to replicate the results of this article are available at \cite{holst_2025_17210834}.

\bibliographystyle{utphys}
\bibliography{references.bib}

\providecommand{\href}[2]{#2}\begingroup\raggedright\begin{thebibliography}{10}

\bibitem{HawkingGravitationallyCollapsedObjects1971}
S.~{Hawking}, ``{Gravitationally collapsed objects of very low mass},''
  \href{https://dx.doi.org/10.1093/mnras/152.1.75}{{\em Mon. Not. Roy. Astron.
  Soc.} {\bfseries 152} (Jan., 1971) 75}.

\bibitem{SasakiPBHsPerspectivesGW2018}
M.~Sasaki, T.~Suyama, T.~Tanaka, and S.~Yokoyama, ``{Primordial black
  holes\textemdash perspectives in gravitational wave astronomy},''
  \href{https://dx.doi.org/10.1088/1361-6382/aaa7b4}{{\em Class. Quantum
  Gravity} {\bfseries 35} no.~6, (2018) 063001},
  \href{https://arxiv.org/abs/1801.05235}{{\ttfamily arXiv:1801.05235
  [astro-ph.CO]}}.

\bibitem{CarrConstraintsPBHs2021}
B.~Carr, K.~Kohri, Y.~Sendouda, and J.~Yokoyama, ``{Constraints on primordial
  black holes},'' \href{https://dx.doi.org/10.1088/1361-6633/ac1e31}{{\em Rept.
  Prog. Phys.} {\bfseries 84} no.~11, (2021) 116902},
  \href{https://arxiv.org/abs/2002.12778}{{\ttfamily arXiv:2002.12778
  [astro-ph.CO]}}.

\bibitem{GreenPBHsDMCandidate2021}
A.~M. Green and B.~J. Kavanagh, ``{Primordial black holes as a dark matter
  candidate},'' \href{https://dx.doi.org/10.1088/1361-6471/abc534}{{\em J.
  Phys. G: Nucl. Part. Phys.} {\bfseries 48} no.~4, (2021) 043001},
  \href{https://arxiv.org/abs/2007.10722}{{\ttfamily arXiv:2007.10722
  [astro-ph.CO]}}.

\bibitem{AcharyaCMBBBNConstraints2020}
S.~K. Acharya and R.~Khatri, ``{CMB and BBN constraints on evaporating
  primordial black holes revisited},''
  \href{https://dx.doi.org/10.1088/1475-7516/2020/06/018}{{\em JCAP} {\bfseries
  06} (2020) 018}, \href{https://arxiv.org/abs/2002.00898}{{\ttfamily
  arXiv:2002.00898 [astro-ph.CO]}}.

\bibitem{HooperDarkRadiationSuperheavy2019}
D.~Hooper, G.~Krnjaic, and S.~D. McDermott, ``{Dark radiation and superheavy
  dark matter from black hole domination},''
  \href{https://dx.doi.org/10.1007/JHEP08(2019)001}{{\em JHEP} {\bfseries 2019}
  no.~8, (2019)}, \href{https://arxiv.org/abs/1905.01301}{{\ttfamily
  arXiv:1905.01301 [hep-ph]}}.

\bibitem{SasakiPBHScenarioGW2016}
M.~Sasaki, T.~Suyama, T.~Tanaka, and S.~Yokoyama, ``{Primordial Black Hole
  Scenario for the Gravitational-Wave Event GW150914},''
  \href{https://dx.doi.org/10.1103/PhysRevLett.117.061101}{{\em PRL} {\bfseries
  117} no.~6, (2016)}, \href{https://arxiv.org/abs/1603.08338}{{\ttfamily
  arXiv:1603.08338 [astro-ph.CO]}}.

\bibitem{RaidalGWsPBHMergers2017}
M.~Raidal, V.~Vaskonen, and H.~Veerm{\"a}e, ``{Gravitational waves from
  primordial black hole mergers},''
  \href{https://dx.doi.org/10.1088/1475-7516/2017/09/037}{{\em JCAP} {\bfseries
  2017} no.~09, (2017) 037--037},
  \href{https://arxiv.org/abs/1707.01480}{{\ttfamily arXiv:1707.01480
  [astro-ph.CO]}}.

\bibitem{ZagoracGUTScalePBHs2019}
J.~L. Zagorac, R.~Easther, and N.~Padmanabhan, ``{GUT-scale primordial black
  holes: mergers and gravitational waves},''
  \href{https://dx.doi.org/10.1088/1475-7516/2019/06/052}{{\em JCAP} {\bfseries
  2019} no.~06, (2019) 052--052},
  \href{https://arxiv.org/abs/1903.05053}{{\ttfamily arXiv:1903.05053
  [astro-ph.CO]}}.

\bibitem{HooperHotGravitonsGWs2020}
D.~Hooper, G.~Krnjaic, J.~March-Russell, S.~D. McDermott, and
  R.~Petrossian-Byrne, ``{Hot Gravitons and Gravitational Waves From Kerr Black
  Holes in the Early Universe},''
  \href{https://arxiv.org/abs/2004.00618}{{\ttfamily arXiv:2004.00618
  [astro-ph.CO]}}.

\bibitem{Shallue:2024hqe}
C.~J. Shallue, J.~B. Mu\~noz, and G.~Z. Krnjaic, ``{Warm Hawking Relics From
  Primordial Black Hole Domination},''
  \href{https://arxiv.org/abs/2406.08535}{{\ttfamily arXiv:2406.08535
  [astro-ph.CO]}}.

\bibitem{Baumann:2007yr}
D.~Baumann, P.~J. Steinhardt, and N.~Turok, ``{Primordial Black Hole
  Baryogenesis},'' \href{https://arxiv.org/abs/hep-th/0703250}{{\ttfamily
  arXiv:hep-th/0703250}}.

\bibitem{MorrisonMelanopogenesisDMalmost2019}
L.~Morrison, S.~Profumo, and Y.~Yu, ``{Melanopogenesis: dark matter of (almost)
  any mass and baryonic matter from the evaporation of primordial black holes
  weighing a ton (or less)},''
  \href{https://dx.doi.org/10.1088/1475-7516/2019/05/005}{{\em JCAP} {\bfseries
  2019} no.~05, (2019) 005--005},
  \href{https://arxiv.org/abs/1812.10606}{{\ttfamily arXiv:1812.10606
  [astro-ph.CO]}}.

\bibitem{HooperGUTBaryogenesisPBHs2021}
D.~Hooper and G.~Krnjaic, ``{GUT baryogenesis with primordial black holes},''
  \href{https://dx.doi.org/10.1103/PhysRevD.103.043504}{{\em PRD} {\bfseries
  103} no.~4, (2021)}, \href{https://arxiv.org/abs/2010.01134}{{\ttfamily
  arXiv:2010.01134 [hep-ph]}}.

\bibitem{Bernal:2022pue}
N.~Bernal, C.~S. Fong, Y.~F. Perez-Gonzalez, and J.~Turner, ``{Rescuing
  high-scale leptogenesis using primordial black holes},''
  \href{https://dx.doi.org/10.1103/PhysRevD.106.035019}{{\em Phys. Rev. D}
  {\bfseries 106} no.~3, (2022) 035019},
  \href{https://arxiv.org/abs/2203.08823}{{\ttfamily arXiv:2203.08823
  [hep-ph]}}.

\bibitem{LennonBHGenesisDM2018}
O.~Lennon, J.~March-Russell, R.~Petrossian-Byrne, and H.~Tillim, ``{Black hole
  genesis of dark matter},''
  \href{https://dx.doi.org/10.1088/1475-7516/2018/04/009}{{\em JCAP} {\bfseries
  2018} no.~04, (2018) 009--009},
  \href{https://arxiv.org/abs/1712.07664}{{\ttfamily arXiv:1712.07664
  [hep-ph]}}.

\bibitem{Cheek:2021odj}
A.~Cheek, L.~Heurtier, Y.~F. Perez-Gonzalez, and J.~Turner, ``{Primordial black
  hole evaporation and dark matter production. I. Solely Hawking radiation},''
  \href{https://dx.doi.org/10.1103/PhysRevD.105.015022}{{\em Phys. Rev. D}
  {\bfseries 105} no.~1, (2022) 015022},
  \href{https://arxiv.org/abs/2107.00013}{{\ttfamily arXiv:2107.00013
  [hep-ph]}}.

\bibitem{AllahverdiFirstThreeSeconds2021}
R.~Allahverdi, M.~A. Amin, {\em et~al.}, ``{The First Three Seconds: a Review
  of Possible Expansion Histories of the Early Universe},''
  \href{https://dx.doi.org/10.21105/astro.2006.16182}{{\em Open J. Astrophys.}
  {\bfseries 4} no.~1, (2021)},
  \href{https://arxiv.org/abs/2006.16182}{{\ttfamily arXiv:2006.16182
  [astro-ph.CO]}}.

\bibitem{PressFormationGalaxiesClusters1974}
W.~H. Press and P.~Schechter, ``{Formation of Galaxies and Clusters of Galaxies
  by Self-Similar Gravitational Condensation},''
  \href{https://dx.doi.org/10.1086/152650}{{\em ApJ} {\bfseries 187} (1974)
  425}.

\bibitem{FishbachAreLIGOsBHs2017}
M.~Fishbach, D.~E. Holz, and B.~Farr, ``{Are LIGO's Black Holes Made from
  Smaller Black Holes?},''
  \href{https://dx.doi.org/10.3847/2041-8213/aa7045}{{\em ApJL} {\bfseries 840}
  no.~2, (2017) L24}, \href{https://arxiv.org/abs/1703.06869}{{\ttfamily
  arXiv:1703.06869 [astro-ph.HE]}}.

\bibitem{DeLucaClusteringEvolutionPBHs2020}
V.~D. Luca, V.~Desjacques, G.~Franciolini, and A.~Riotto, ``{The clustering
  evolution of primordial black holes},''
  \href{https://dx.doi.org/10.1088/1475-7516/2020/11/028}{{\em JCAP} {\bfseries
  2020} no.~11, (2020) 028--028},
  \href{https://arxiv.org/abs/2009.04731}{{\ttfamily arXiv:2009.04731
  [astro-ph.CO]}}.

\bibitem{FrancioliniPBHMergersThree2022}
G.~Franciolini, K.~Kritos, E.~Berti, and J.~Silk, ``{Primordial black hole
  mergers from three-body interactions},''
  \href{https://dx.doi.org/10.1103/PhysRevD.106.083529}{{\em PRD} {\bfseries
  106} no.~8, (2022)}, \href{https://arxiv.org/abs/2205.15340}{{\ttfamily
  arXiv:2205.15340 [astro-ph.CO]}}.

\bibitem{DelosStructureFormationPBHs2024}
M.~S. Delos, A.~Rantala, S.~Young, and F.~Schmidt, ``{Structure formation with
  primordial black holes: collisional dynamics, binaries, and gravitational
  waves},'' \href{https://arxiv.org/abs/2410.01876}{{\ttfamily arXiv:2410.01876
  [astro-ph.CO]}}.

\bibitem{NakamuraGWsCoalescingBH1997}
T.~Nakamura, M.~Sasaki, T.~Tanaka, and K.~S. Thorne, ``{Gravitational Waves
  from Coalescing Black Hole MACHO Binaries},''
  \href{https://dx.doi.org/10.1086/310886}{{\em ApJ} {\bfseries 487} no.~2,
  (1997) L139--L142}, \href{https://arxiv.org/abs/astro-ph/9708060}{{\ttfamily
  arXiv:astro-ph/9708060}}.

\bibitem{IokaBHBinaryFormation1998}
K.~Ioka, T.~Chiba, T.~Tanaka, and T.~Nakamura, ``{Black hole binary formation
  in the expanding universe: Three body problem approximation},''
  \href{https://dx.doi.org/10.1103/PhysRevD.58.063003}{{\em PRD} {\bfseries 58}
  no.~6, (1998)}, \href{https://arxiv.org/abs/astro-ph/9807018}{{\ttfamily
  arXiv:astro-ph/9807018}}.

\bibitem{AliHaimoudMergerRatePBH2017}
Y.~Ali-Ha{\"\i}moud, E.~D. Kovetz, and M.~Kamionkowski, ``{Merger rate of
  primordial black-hole binaries},''
  \href{https://dx.doi.org/10.1103/PhysRevD.96.123523}{{\em PRD} {\bfseries 96}
  no.~12, (2017)}, \href{https://arxiv.org/abs/1709.06576}{{\ttfamily
  arXiv:1709.06576 [astro-ph.CO]}}.

\bibitem{RaidalFormationEvolutionPBH2019}
M.~Raidal, C.~Spethmann, V.~Vaskonen, and H.~Veerm{\"a}e, ``{Formation and
  evolution of primordial black hole binaries in the early universe},''
  \href{https://dx.doi.org/10.1088/1475-7516/2019/02/018}{{\em JCAP} {\bfseries
  2019} no.~02, (2019) 018--018},
  \href{https://arxiv.org/abs/1812.01930}{{\ttfamily arXiv:1812.01930
  [astro-ph.CO]}}.

\bibitem{DeLucaHeavyPBHsStrongly2023}
V.~De~Luca, G.~Franciolini, and A.~Riotto, ``{Heavy Primordial Black Holes from
  Strongly Clustered Light Black Holes},''
  \href{https://dx.doi.org/10.1103/PhysRevLett.130.171401}{{\em PRL} {\bfseries
  130} no.~17, (2023)}, \href{https://arxiv.org/abs/2210.14171}{{\ttfamily
  arXiv:2210.14171 [astro-ph.CO]}}.

\bibitem{ChisholmClusteringPBHsII2011}
J.~R. Chisholm, ``{Clustering of Primordial Black Holes. II. Evolution of Bound
  Systems},'' \href{https://dx.doi.org/10.1103/PhysRevD.84.124031}{{\em Phys.
  Rev. D} {\bfseries 84} (2011) 124031},
  \href{https://arxiv.org/abs/1110.4402}{{\ttfamily arXiv:1110.4402
  [astro-ph.CO]}}.

\bibitem{KimPBHReformationEarly2024}
T.~Kim and P.~Lu, ``{Primordial Black Hole Reformation in the Early
  Universe},'' \href{https://arxiv.org/abs/2411.07469}{{\ttfamily
  arXiv:2411.07469 [astro-ph.CO]}}.

\bibitem{XuDynamicsMassiveBHs1994}
G.~Xu and J.~P. Ostriker, ``{Dynamics of massive black holes as a possible
  candidate of Galactic dark matter},''
  \href{https://dx.doi.org/10.1086/174987}{{\em ApJ} {\bfseries 437} (1994)
  184}.

\bibitem{TanakaAssemblySupermassiveBHs2009}
T.~Tanaka and Z.~Haiman, ``{The Assembly of Supermassive Black Holes at High
  Redshifts},'' \href{https://dx.doi.org/10.1088/0004-637X/696/2/1798}{{\em
  Astrophys. J.} {\bfseries 696} (2009) 1798--1822},
  \href{https://arxiv.org/abs/0807.4702}{{\ttfamily arXiv:0807.4702
  [astro-ph]}}.

\bibitem{InayoshiAssemblyFirstMassive2020}
K.~Inayoshi, E.~Visbal, and Z.~Haiman, ``{The Assembly of the First Massive
  Black Holes},''
  \href{https://dx.doi.org/10.1146/annurev-astro-120419-014455}{{\em Ann. Rev.
  Astron. Astrophys.} {\bfseries 58} (2020) 27--97},
  \href{https://arxiv.org/abs/1911.05791}{{\ttfamily arXiv:1911.05791
  [astro-ph.GA]}}.

\bibitem{HawkingBHExplosions1974}
S.~W. Hawking, ``{Black hole explosions?},''
  \href{https://dx.doi.org/10.1038/248030a0}{{\em Nature} {\bfseries 248}
  no.~5443, (1974) 30--31}.

\bibitem{HawkingParticleCreationBHs1975}
S.~W. Hawking, ``{Particle Creation by Black Holes},''
  \href{https://dx.doi.org/10.1007/BF02345020}{{\em Commun. Math. Phys.}
  {\bfseries 43} (1975) 199--220}.

\bibitem{MacGibbonQuarkGluonJet1990}
J.~H. MacGibbon and B.~R. Webber, ``{Quark- and gluon-jet emission from
  primordial black holes: The instantaneous spectra},''
  \href{https://dx.doi.org/10.1103/PhysRevD.41.3052}{{\em Phys. Rev. D}
  {\bfseries 41} (1990) 3052--3079}.

\bibitem{MacGibbonQuarkGluonJet1991}
J.~H. MacGibbon, ``{Quark and gluon jet emission from primordial black holes.
  2. The Lifetime emission},''
  \href{https://dx.doi.org/10.1103/PhysRevD.44.376}{{\em Phys. Rev. D}
  {\bfseries 44} (1991) 376--392}.

\bibitem{LiuPBHProductionDuring2022}
J.~Liu, L.~Bian, R.-G. Cai, Z.-K. Guo, and S.-J. Wang, ``{Primordial black hole
  production during first-order phase transitions},''
  \href{https://dx.doi.org/10.1103/PhysRevD.105.L021303}{{\em PRD} {\bfseries
  105} no.~2, (2022)}, \href{https://arxiv.org/abs/2106.05637}{{\ttfamily
  arXiv:2106.05637 [astro-ph.CO]}}.

\bibitem{RubinPBHsNonequilibriumSecond2000}
S.~G. Rubin, M.~Y. Khlopov, and A.~S. Sakharov, ``{Primordial black holes from
  nonequilibrium second order phase transition},'' {\em Grav. Cosmol.}
  {\bfseries 6} (2000) 51--58,
  \href{https://arxiv.org/abs/hep-ph/0005271}{{\ttfamily
  arXiv:hep-ph/0005271}}.

\bibitem{DunskyPBHsAxionDomain2024}
D.~I. Dunsky and M.~Kongsore, ``{Primordial black holes from axion domain wall
  collapse},'' \href{https://dx.doi.org/10.1007/JHEP06(2024)198}{{\em JHEP}
  {\bfseries 06} (2024) 198},
  \href{https://arxiv.org/abs/2402.03426}{{\ttfamily arXiv:2402.03426
  [hep-ph]}}.

\bibitem{CotnerPBHsSupersymmetryEarly2017}
E.~Cotner and A.~Kusenko, ``{Primordial Black Holes from Supersymmetry in the
  Early Universe},''
  \href{https://dx.doi.org/10.1103/PhysRevLett.119.031103}{{\em PRL} {\bfseries
  119} no.~3, (2017)}, \href{https://arxiv.org/abs/1612.02529}{{\ttfamily
  arXiv:1612.02529 [astro-ph.CO]}}.

\bibitem{Lu:2024xnb}
Y.~Lu, Z.~S.~C. Picker, S.~Profumo, and A.~Kusenko, ``{Black Holes from Fermi
  Ball Collapse},'' \href{https://arxiv.org/abs/2411.17074}{{\ttfamily
  arXiv:2411.17074 [astro-ph.CO]}}.

\bibitem{KibbleTopologyCosmicDomains1976}
T.~W.~B. Kibble, ``{Topology of cosmic domains and strings},''
  \href{https://dx.doi.org/10.1088/0305-4470/9/8/029}{{\em J. Phys. A: Math.
  Gen.} {\bfseries 9} no.~8, (1976) 1387--1398}.

\bibitem{VilenkinCosmicStringsPBHs2018}
A.~Vilenkin, Y.~Levin, and A.~Gruzinov, ``{Cosmic strings and primordial black
  holes},'' \href{https://dx.doi.org/10.1088/1475-7516/2018/11/008}{{\em JCAP}
  {\bfseries 2018} no.~11, (2018) 008--008},
  \href{https://arxiv.org/abs/1808.00670}{{\ttfamily arXiv:1808.00670
  [astro-ph.CO]}}.

\bibitem{FloresStructureFormationAfter2023}
M.~M. Flores, Y.~Lu, and A.~Kusenko, ``{Structure formation after reheating:
  Supermassive primordial black holes and Fermi ball dark matter},''
  \href{https://dx.doi.org/10.1103/PhysRevD.108.123511}{{\em Phys. Rev. D}
  {\bfseries 108} no.~12, (2023) 123511},
  \href{https://arxiv.org/abs/2308.09094}{{\ttfamily arXiv:2308.09094
  [astro-ph.CO]}}.

\bibitem{BramanteDissipativeDarkCosmology2024}
J.~Bramante, C.~V. Cappiello, M.~Diamond, J.~L. Kim, Q.~Liu, and A.~C. Vincent,
  ``{Dissipative dark cosmology: From early matter dominance to delayed compact
  objects},'' \href{https://dx.doi.org/10.1103/PhysRevD.110.043041}{{\em Phys.
  Rev. D} {\bfseries 110} no.~4, (2024) 043041},
  \href{https://arxiv.org/abs/2405.04575}{{\ttfamily arXiv:2405.04575
  [hep-ph]}}.

\bibitem{RalegankarGravothermalizingPBHsBoson2024}
P.~Ralegankar, D.~Perri, and T.~Kobayashi, ``{Gravothermalizing into primordial
  black holes, boson stars, and cannibal stars},''
  \href{https://arxiv.org/abs/2410.18948}{{\ttfamily arXiv:2410.18948
  [astro-ph.CO]}}.

\bibitem{ParticleDataGroupReviewParticlePhysics2022}
{\bfseries Particle Data Group} Collaboration, R.~L. Workman, V.~D. Burkert,
  {\em et~al.}, ``{Review of Particle Physics},''
  \href{https://dx.doi.org/10.1093/ptep/ptac097}{{\em PTEP} {\bfseries 2022}
  no.~8, (2022)}.

\bibitem{DesjacquesSpatialClusteringPBHs2018}
V.~Desjacques and A.~Riotto, ``{Spatial clustering of primordial black
  holes},'' \href{https://dx.doi.org/10.1103/PhysRevD.98.123533}{{\em PRD}
  {\bfseries 98} no.~12, (2018)},
  \href{https://arxiv.org/abs/1806.10414}{{\ttfamily arXiv:1806.10414
  [astro-ph.CO]}}.

\bibitem{InmanEarlyStructureFormation2019}
D.~Inman and Y.~Ali-Ha{\"\i}moud, ``{Early structure formation in primordial
  black hole cosmologies},''
  \href{https://dx.doi.org/10.1103/PhysRevD.100.083528}{{\em PRD} {\bfseries
  100} no.~8, (2019)}, \href{https://arxiv.org/abs/1907.08129}{{\ttfamily
  arXiv:1907.08129 [astro-ph.CO]}}.

\bibitem{DomenechExploringEvaporatingPBHs2021}
G.~Dom{\`e}nech, V.~Takhistov, and M.~Sasaki, ``{Exploring evaporating
  primordial black holes with gravitational waves},''
  \href{https://dx.doi.org/10.1016/j.physletb.2021.136722}{{\em Phys. Lett. B}
  {\bfseries 823} (2021) 136722},
  \href{https://arxiv.org/abs/2105.06816}{{\ttfamily arXiv:2105.06816
  [astro-ph.CO]}}.

\bibitem{AuclairSmallScaleClustering2024}
P.~Auclair and B.~Blachier, ``{Small-scale clustering of Primordial Black
  Holes: cloud-in-cloud and exclusion effects},''
  \href{https://arxiv.org/abs/2402.00600}{{\ttfamily arXiv:2402.00600
  [astro-ph.CO]}}.

\bibitem{DodelsonModernCosmology}
S.~Dodelson, {\em {Modern Cosmology}}.
\newblock Academic Press, Amsterdam, 2003.

\bibitem{CoorayHaloModelsLarge2002}
A.~Cooray and R.~K. Sheth, ``{Halo Models of Large Scale Structure},''
  \href{https://dx.doi.org/10.1016/S0370-1573(02)00276-4}{{\em Phys. Rept.}
  {\bfseries 372} (2002) 1--129},
  \href{https://arxiv.org/abs/astro-ph/0206508}{{\ttfamily
  arXiv:astro-ph/0206508}}.

\bibitem{MoBoschWhite}
H.~{Mo}, F.~C. {van den Bosch}, and S.~{White}, {\em Galaxy Formation and
  Evolution}.
\newblock Cambridge University Press, 2010.

\bibitem{BinneyTremaine}
J.~Binney and S.~Tremaine, {\em Galactic Dynamics}.
\newblock Princeton University Press, 1988.

\bibitem{BlancoAnnihilationSignaturesHidden2019}
C.~Blanco, M.~S. Delos, A.~L. Erickcek, and D.~Hooper, ``{Annihilation
  signatures of hidden sector dark matter within early-forming microhalos},''
  \href{https://dx.doi.org/10.1103/PhysRevD.100.103010}{{\em PRD} {\bfseries
  100} no.~10, (2019)}, \href{https://arxiv.org/abs/1906.00010}{{\ttfamily
  arXiv:1906.00010 [astro-ph.CO]}}.

\bibitem{XiaoSimulationsAxionMinihalos2021}
H.~Xiao, I.~Williams, and M.~McQuinn, ``{Simulations of axion minihalos},''
  \href{https://dx.doi.org/10.1103/PhysRevD.104.023515}{{\em PRD} {\bfseries
  104} no.~2, (2021)}, \href{https://arxiv.org/abs/2101.04177}{{\ttfamily
  arXiv:2101.04177 [astro-ph.CO]}}.

\bibitem{CholisOrbitalEccentricitiesPBH2016}
I.~Cholis, E.~D. Kovetz, Y.~Ali-Ha{\"\i}moud, S.~Bird, M.~Kamionkowski, J.~B.
  Mu{\~n}oz, and A.~Raccanelli, ``{Orbital eccentricities in primordial black
  hole binaries},'' \href{https://dx.doi.org/10.1103/PhysRevD.94.084013}{{\em
  PRD} {\bfseries 94} no.~8, (2016)},
  \href{https://arxiv.org/abs/1606.07437}{{\ttfamily arXiv:1606.07437
  [astro-ph.HE]}}.

\bibitem{LightmanDynamicalEvolutionGlobular1978}
A.~P. Lightman and S.~L. Shapiro, ``{The dynamical evolution of globular
  clusters},'' \href{https://dx.doi.org/10.1103/RevModPhys.50.437}{{\em Rev.
  Mod. Phys.} {\bfseries 50} no.~2, (1978) 437--481}.

\bibitem{MisnerThorneWheeler}
C.~W. Misner, K.~S. Thorne, and J.~A. Wheeler, {\em {Gravitation}}.
\newblock W. H. Freeman, San Francisco, 1973.

\bibitem{TurnerGravitationalRadiationPoint1977}
M.~Turner, ``{Gravitational radiation from point-masses in unbound orbits -
  Newtonian results},'' \href{https://dx.doi.org/10.1086/155501}{{\em ApJ}
  {\bfseries 216} (1977) 610}.

\bibitem{QuinlanDynamicalEvolutionDense1989}
G.~D. Quinlan and S.~L. Shapiro, ``{Dynamical evolution of dense clusters of
  compact stars},'' \href{https://dx.doi.org/10.1086/167745}{{\em ApJ}
  {\bfseries 343} (1989) 725}.

\bibitem{LeeNBodyEvolution1993}
M.~H. Lee, ``{N-Body Evolution of Dense Clusters of Compact Stars},''
  \href{https://dx.doi.org/10.1086/173378}{{\em ApJ} {\bfseries 418} (1993)
  147}.

\bibitem{OLearyGWsScatteringStellar2009}
R.~M. O’Leary, B.~Kocsis, and A.~Loeb, ``{Gravitational waves from scattering
  of stellar-mass black holes in galactic nuclei},''
  \href{https://dx.doi.org/10.1111/j.1365-2966.2009.14653.x}{{\em MNRAS}
  {\bfseries 395} no.~4, (2009) 2127--2146},
  \href{https://arxiv.org/abs/0807.2638}{{\ttfamily arXiv:0807.2638
  [astro-ph]}}.

\bibitem{PetersGravitationalRadiationMotion1964}
P.~C. Peters, ``{Gravitational Radiation and the Motion of Two Point Masses},''
  \href{https://dx.doi.org/10.1103/PhysRev.136.B1224}{{\em Phys. Rev.}
  {\bfseries 136} no.~4B, (1964) B1224--B1232}.

\bibitem{BirdDidLIGODetect2016}
S.~Bird, I.~Cholis, J.~B. Mu\~noz, Y.~Ali-Ha\"\i{}moud, M.~Kamionkowski, E.~D.
  Kovetz, A.~Raccanelli, and A.~G. Riess, ``{Did LIGO detect dark matter?},''
  \href{https://dx.doi.org/10.1103/PhysRevLett.116.201301}{{\em Phys. Rev.
  Lett.} {\bfseries 116} no.~20, (2016) 201301},
  \href{https://arxiv.org/abs/1603.00464}{{\ttfamily arXiv:1603.00464
  [astro-ph.CO]}}.

\bibitem{AfshordiPBHsDMPower2003}
N.~Afshordi, P.~McDonald, and D.~N. Spergel, ``{Primordial Black Holes as Dark
  Matter: The Power Spectrum and Evaporation of Early Structures},''
  \href{https://dx.doi.org/10.1086/378763}{{\em ApJ} {\bfseries 594} no.~2,
  (2003) L71--L74}, \href{https://arxiv.org/abs/astro-ph/0302035}{{\ttfamily
  arXiv:astro-ph/0302035}}.

\bibitem{CeloriaLectureNotesBH2018}
M.~Celoria, R.~Oliveri, A.~Sesana, and M.~Mapelli, ``{Lecture notes on black
  hole binary astrophysics},''
  \href{https://arxiv.org/abs/1807.11489}{{\ttfamily arXiv:1807.11489
  [astro-ph.GA]}}.

\bibitem{AmbartsumianDynamicsOpenClusters1938}
V.~A. {Ambartsumian}, ``{On the dynamics of open clusters},'' {\em Ann.
  Leningrad State Univ.} {\bfseries 22} (1938) 19.

\bibitem{SpitzerStabilityIsolatedClusters1940}
L.~Spitzer and H.~Shapley, ``{The Stability of Isolated Clusters},''
  \href{https://dx.doi.org/10.1093/mnras/100.5.396}{{\em MNRAS} {\bfseries 100}
  no.~5, (1940) 396--413}.

\bibitem{ChandrasekharDynamicalFrictionIII1943}
S.~Chandrasekhar, ``{Dynamical Friction. III. a More Exact Theory of the Rate
  of Escape of Stars from Clusters.},''
  \href{https://dx.doi.org/10.1086/144544}{{\em ApJ} {\bfseries 98} (1943) 54}.

\bibitem{LIGOScientificVirgoObservationGWsBinary2016}
{\bfseries LIGO Scientific, Virgo} Collaboration, B.~P. Abbott {\em et~al.},
  ``{Observation of Gravitational Waves from a Binary Black Hole Merger},''
  \href{https://dx.doi.org/10.1103/PhysRevLett.116.061102}{{\em Phys. Rev.
  Lett.} {\bfseries 116} no.~6, (2016) 061102},
  \href{https://arxiv.org/abs/1602.03837}{{\ttfamily arXiv:1602.03837
  [gr-qc]}}.

\bibitem{LyndenBellGravoThermalCatastrophe1968}
D.~Lynden-Bell, R.~Wood, and A.~Royal, ``{The Gravo-Thermal Catastrophe in
  Isothermal Spheres and the Onset of Red-Giant Structure for Stellar
  Systems},'' \href{https://dx.doi.org/10.1093/mnras/138.4.495}{{\em Mon. Not.
  Roy. Astron. Soc.} {\bfseries 138} no.~4, (1968) 495--525}.

\bibitem{BalbergSelfinteractingDMHalos2002}
S.~Balberg, S.~L. Shapiro, and S.~Inagaki, ``{Selfinteracting dark matter halos
  and the gravothermal catastrophe},''
  \href{https://dx.doi.org/10.1086/339038}{{\em Astrophys. J.} {\bfseries 568}
  (2002) 475--487}, \href{https://arxiv.org/abs/astro-ph/0110561}{{\ttfamily
  arXiv:astro-ph/0110561}}.

\bibitem{Martin:2013tda}
J.~Martin, C.~Ringeval, and V.~Vennin, ``{Encyclop{\ae}dia Inflationaris}:
  {Opiparous Edition},''
  \href{https://dx.doi.org/10.1016/j.dark.2024.101653}{{\em Phys. Dark Univ.}
  {\bfseries 5-6} (2014) 75--235},
  \href{https://arxiv.org/abs/1303.3787}{{\ttfamily arXiv:1303.3787
  [astro-ph.CO]}}.

\bibitem{CarrNewCosmologicalConstraints2010}
B.~J. Carr, K.~Kohri, Y.~Sendouda, and J.~Yokoyama, ``{New cosmological
  constraints on primordial black holes},''
  \href{https://dx.doi.org/10.1103/PhysRevD.81.104019}{{\em Phys. Rev. D}
  {\bfseries 81} (2010) 104019},
  \href{https://arxiv.org/abs/0912.5297}{{\ttfamily arXiv:0912.5297
  [astro-ph.CO]}}.

\bibitem{Boudaud:2018hqb}
M.~Boudaud and M.~Cirelli, ``{Voyager 1 $e^\pm$ Further Constrain Primordial
  Black Holes as Dark Matter},''
  \href{https://dx.doi.org/10.1103/PhysRevLett.122.041104}{{\em Phys. Rev.
  Lett.} {\bfseries 122} no.~4, (2019) 041104},
  \href{https://arxiv.org/abs/1807.03075}{{\ttfamily arXiv:1807.03075
  [astro-ph.HE]}}.

\bibitem{BellomoPBHsDMConverting2018}
N.~Bellomo, J.~L. Bernal, A.~Raccanelli, and L.~Verde, ``{Primordial black
  holes as dark matter: converting constraints from monochromatic to extended
  mass distributions},''
  \href{https://dx.doi.org/10.1088/1475-7516/2018/01/004}{{\em JCAP} {\bfseries
  2018} no.~01, (2018) 004--004},
  \href{https://arxiv.org/abs/1709.07467}{{\ttfamily arXiv:1709.07467
  [astro-ph.CO]}}.

\bibitem{CMBS4ScienceBook2016}
{\bfseries CMB-S4} Collaboration, K.~N. Abazajian {\em et~al.}, ``{CMB-S4
  Science Book, First Edition},''
  \href{https://arxiv.org/abs/1610.02743}{{\ttfamily arXiv:1610.02743
  [astro-ph.CO]}}.

\bibitem{FlanaganMeasuringGWsBinary1998}
E.~E. Flanagan and S.~A. Hughes, ``{Measuring gravitational waves from binary
  black hole coalescences: 1. Signal-to-noise for inspiral, merger, and
  ringdown},'' \href{https://dx.doi.org/10.1103/PhysRevD.57.4535}{{\em Phys.
  Rev. D} {\bfseries 57} (1998) 4535--4565},
  \href{https://arxiv.org/abs/gr-qc/9701039}{{\ttfamily arXiv:gr-qc/9701039}}.

\bibitem{AjithInspiralMergerRingdown2011}
P.~Ajith {\em et~al.}, ``{Inspiral-merger-ringdown waveforms for black-hole
  binaries with non-precessing spins},''
  \href{https://dx.doi.org/10.1103/PhysRevLett.106.241101}{{\em Phys. Rev.
  Lett.} {\bfseries 106} (2011) 241101},
  \href{https://arxiv.org/abs/0909.2867}{{\ttfamily arXiv:0909.2867 [gr-qc]}}.

\bibitem{SetoPossibilityDirectMeasurement2001}
N.~Seto, S.~Kawamura, and T.~Nakamura, ``{Possibility of direct measurement of
  the acceleration of the universe using 0.1-Hz band laser interferometer
  gravitational wave antenna in space},''
  \href{https://dx.doi.org/10.1103/PhysRevLett.87.221103}{{\em Phys. Rev.
  Lett.} {\bfseries 87} (2001) 221103},
  \href{https://arxiv.org/abs/astro-ph/0108011}{{\ttfamily
  arXiv:astro-ph/0108011}}.

\bibitem{YagiDetectorConfigurationDECIGOBBO2011}
K.~Yagi and N.~Seto, ``{Detector configuration of DECIGO/BBO and identification
  of cosmological neutron-star binaries},''
  \href{https://dx.doi.org/10.1103/PhysRevD.83.044011}{{\em Phys. Rev. D}
  {\bfseries 83} (2011) 044011},
  \href{https://arxiv.org/abs/1101.3940}{{\ttfamily arXiv:1101.3940
  [astro-ph.CO]}}.

\bibitem{CheekRedshiftEffectsParticle2022}
A.~Cheek, L.~Heurtier, Y.~F. Perez-Gonzalez, and J.~Turner, ``{Redshift effects
  in particle production from Kerr primordial black holes},''
  \href{https://dx.doi.org/10.1103/PhysRevD.106.103012}{{\em PRD} {\bfseries
  106} no.~10, (2022)}, \href{https://arxiv.org/abs/2207.09462}{{\ttfamily
  arXiv:2207.09462 [astro-ph.CO]}}.

\bibitem{CheekEvaporationPBHsEarly2023}
A.~Cheek, L.~Heurtier, Y.~F. Perez-Gonzalez, and J.~Turner, ``{Evaporation of
  primordial black holes in the early Universe: Mass and spin distributions},''
  \href{https://dx.doi.org/10.1103/PhysRevD.108.015005}{{\em PRD} {\bfseries
  108} no.~1, (2023)}, \href{https://arxiv.org/abs/2212.03878}{{\ttfamily
  arXiv:2212.03878 [hep-ph]}}.

\bibitem{PapanikolaouGWsUniverseFilled2021}
T.~Papanikolaou, V.~Vennin, and D.~Langlois, ``{Gravitational waves from a
  universe filled with primordial black holes},''
  \href{https://dx.doi.org/10.1088/1475-7516/2021/03/053}{{\em JCAP} {\bfseries
  2021} no.~03, (2021) 053}, \href{https://arxiv.org/abs/2010.11573}{{\ttfamily
  arXiv:2010.11573 [astro-ph.CO]}}.

\bibitem{BaumannGWSpectrumInduced2007}
D.~Baumann, P.~Steinhardt, K.~Takahashi, and K.~Ichiki, ``{Gravitational wave
  spectrum induced by primordial scalar perturbations},''
  \href{https://dx.doi.org/10.1103/PhysRevD.76.084019}{{\em PRD} {\bfseries 76}
  no.~8, (2007)}, \href{https://arxiv.org/abs/hep-th/0703290}{{\ttfamily
  arXiv:hep-th/0703290}}.

\bibitem{DomenechGWConstraintsPBH2021}
G.~Dom{\`e}nech, C.~Lin, and M.~Sasaki, ``{Gravitational wave constraints on
  the primordial black hole dominated early universe},''
  \href{https://dx.doi.org/10.1088/1475-7516/2021/04/062}{{\em JCAP} {\bfseries
  2021} no.~04, (2021) 062}, \href{https://arxiv.org/abs/2012.08151}{{\ttfamily
  arXiv:2012.08151 [gr-qc]}}.

\bibitem{EggemeierStochasticGWsPostinflationary2023}
B.~Eggemeier, J.~C. Niemeyer, K.~Jedamzik, and R.~Easther, ``{Stochastic
  gravitational waves from postinflationary structure formation},''
  \href{https://dx.doi.org/10.1103/PhysRevD.107.043503}{{\em PRD} {\bfseries
  107} no.~4, (2023)}, \href{https://arxiv.org/abs/2212.00425}{{\ttfamily
  arXiv:2212.00425 [astro-ph.CO]}}.

\bibitem{EbadiGWsStochasticScalar2024}
R.~Ebadi, S.~Kumar, A.~McCune, H.~Tai, and L.-T. Wang, ``{Gravitational waves
  from stochastic scalar fluctuations},''
  \href{https://dx.doi.org/10.1103/PhysRevD.109.083519}{{\em Phys. Rev. D}
  {\bfseries 109} no.~8, (2024) 083519},
  \href{https://arxiv.org/abs/2307.01248}{{\ttfamily arXiv:2307.01248
  [astro-ph.CO]}}.

\bibitem{DomenechScalarInducedGWs2021}
G.~Domenech, ``{Scalar Induced Gravitational Waves Review},''
  \href{https://dx.doi.org/10.3390/universe7110398}{{\em Universe} {\bfseries
  7} no.~11, (2021) 398}, \href{https://arxiv.org/abs/2109.01398}{{\ttfamily
  arXiv:2109.01398 [gr-qc]}}.

\bibitem{FernandezStochasticGWsEarly2024}
N.~Fernandez, J.~W. Foster, B.~Lillard, and J.~Shelton, ``{Stochastic
  Gravitational Waves from Early Structure Formation},''
  \href{https://dx.doi.org/10.1103/PhysRevLett.133.111002}{{\em Phys. Rev.
  Lett.} {\bfseries 133} no.~11, (2024) 111002},
  \href{https://arxiv.org/abs/2312.12499}{{\ttfamily arXiv:2312.12499
  [astro-ph.CO]}}.

\bibitem{InomataEnhancementGWsInduced2019}
K.~Inomata, K.~Kohri, T.~Nakama, and T.~Terada, ``{Enhancement of gravitational
  waves induced by scalar perturbations due to a sudden transition from an
  early matter era to the radiation era},''
  \href{https://dx.doi.org/10.1103/PhysRevD.100.043532}{{\em PRD} {\bfseries
  100} no.~4, (2019)}, \href{https://arxiv.org/abs/1904.12879}{{\ttfamily
  arXiv:1904.12879 [astro-ph.CO]}}.

\bibitem{InomataGWProductionRight2020}
K.~Inomata, M.~Kawasaki, K.~Mukaida, T.~Terada, and T.~T. Yanagida,
  ``{Gravitational wave production right after a primordial black hole
  evaporation},'' \href{https://dx.doi.org/10.1103/PhysRevD.101.123533}{{\em
  PRD} {\bfseries 101} no.~12, (2020)},
  \href{https://arxiv.org/abs/2003.10455}{{\ttfamily arXiv:2003.10455
  [astro-ph.CO]}}.

\bibitem{holst_2025_17210834}
I.~Holst, G.~Krnjaic, and H.~Xiao, ``{Code and Data for Clustering and Runaway
  Merging in a Primordial Black Hole Dominated Universe}.'' 2025.
\newblock \url{https://doi.org/10.5281/zenodo.17210834}.

\end{thebibliography}\endgroup

\onecolumngrid
\appendix

\section{Shot Noise Power Spectrum}\label{app:shot_noise}

\newcommand{\vx}{\mathbf{x}}
\newcommand{\vk}{\mathbf{k}}

Here we present a full derivation of the often-quoted result for the power spectrum of fluctuations due to a distribution of point sources with no inherent clustering. We consider a randomly distributed set of black holes of mass $m$. Treating them as point masses at positions $\vx_i$, their number density is a sum of delta functions
\begin{equation}
    n_\BH(\vx) = \sum_i \delta_D^{(3)}(\vx - \vx_i).
\end{equation}
If the universe is dominated by black holes, then the average BH number density is $\bar{n}_\BH = \bar{\rho} / m$. We define the black hole overdensity
\begin{equation}
    \delta_\BH(\vx) = \frac{n_\BH(\vx) - \bar{n}_\BH}{\bar{n}_\BH} = \frac{1}{\bar{n}_\BH}\sum_i \delta_D^{(3)}(\vx - \vx_i) - 1.
\end{equation}
The Fourier transform of the overdensity is then a sum of plane waves with the constant component subtracted:
\begin{equation}
\begin{split}
    \delta_\BH(\vk) & = \int d^3\vx\, \delta_\BH(\vx) e^{-i \vk \cdot \vx} \\
        & = \int d^3\vx\, \left(\frac{1}{\bar{n}_\BH}\sum_i \delta_D^{(3)}(\vx - \vx_i) - 1\right) e^{-i \vk \cdot \vx} \\
        & = \frac{1}{\bar{n}_\BH} \sum_i \int d^3\vx\, \delta_D^{(3)}(\vx - \vx_i) e^{-i \vk \cdot \vx} - \int d^3\vx\, e^{-i \vk \cdot \vx} \\
        & = \frac{1}{\bar{n}_\BH} \sum_i e^{-i \vk \cdot \vx_i} - (2\pi)^3\delta_D^{(3)}(\vk) \\
\end{split}
\end{equation}

The power spectrum of black hole fluctuations is defined
\begin{equation}
    \langle \delta_\BH(\vk) \delta_\BH^{*}(\vk') \rangle = (2\pi)^3 \delta_D^{(3)}(\vk-\vk') P_\BH(k),
\end{equation}
and we combine this with the Fourier transform of the overdensity above, carrying through the ensemble averages on the set of $\{\vx_i\}$:
\begin{equation}
\begin{split}
    (2\pi)^3 \delta_D^{(3)}(\vk-\vk') P_\BH(k) & = \left\langle \left( \frac{1}{\bar{n}_\BH} \sum_i e^{-i \vk \cdot \vx_i} - (2\pi)^3\delta_D^{(3)}(\vk) \right) \left( \frac{1}{\bar{n}_\BH} \sum_j e^{i \vk' \cdot \vx_j} - (2\pi)^3\delta_D^{(3)}(\vk') \right) \right\rangle \\
        & = \frac{1}{\bar{n}_\BH^2} \sum_i \sum_j \left\langle e^{-i \vk \cdot \vx_i} e^{i \vk' \cdot \vx_j} \right\rangle \\
        & ~~~~~~~~~~ - \frac{1}{\bar{n}_\BH} \left(  \sum_i \left\langle e^{-i \vk \cdot \vx_i} \right\rangle (2\pi)^3\delta_D^{(3)}(\vk') +  \sum_j \left\langle e^{i \vk' \cdot \vx_j} \right\rangle (2\pi)^3\delta_D^{(3)}(\vk) \right) \\
        & ~~~~~~~~~~ + (2\pi)^6 \delta_D^{(3)}(\vk) \delta_D^{(3)}(\vk')
\end{split}
\end{equation}
In the first term we can separate the cases where $i = j$ and $i \neq j$
\begin{equation}
\begin{split}
    (2\pi)^3 \delta_D^{(3)}(\vk-\vk') P_\BH(k) & = \frac{1}{\bar{n}_\BH^2} \left( \sum_{i \neq j} \left\langle e^{-i \vk \cdot \vx_i} e^{i \vk' \cdot \vx_j} \right\rangle + \sum_i \left\langle e^{-i (\vk-\vk') \cdot \vx_i} \right\rangle \right)\\
    & ~~~~~~~~~~ - \frac{1}{\bar{n}_\BH} \left(  \sum_i \left\langle e^{-i \vk \cdot \vx_i} \right\rangle (2\pi)^3\delta_D^{(3)}(\vk') +  \sum_j \left\langle e^{i \vk' \cdot \vx_j} \right\rangle (2\pi)^3\delta_D^{(3)}(\vk) \right) \\
    & ~~~~~~~~~~ + (2\pi)^6 \delta_D^{(3)}(\vk) \delta_D^{(3)}(\vk')
\end{split}
\end{equation}
The ensemble averages of plane waves can be understood as volume integrals weighted by a plane wave, which is also a Fourier transform, giving a delta function in $k$. We introduce a factor of $V$ for the total volume in the denominator.
\begin{equation}
    \left\langle e^{-i \vk \cdot \vx_i} \right\rangle = \frac{\int d^3\vx_i \, e^{-i \vk \cdot \vx_i}}{\int d^3\vx_i} = \frac{(2\pi)^3\delta_D^{(3)}(\vk)}{V}
\end{equation}
\begin{equation}
    \left\langle e^{-i \vk \cdot \vx_i} e^{i \vk' \cdot \vx_j} \right\rangle = \frac{\int d^3\vx_i \, d^3\vx_j \, e^{-i \vk \cdot \vx_i} e^{i \vk' \cdot \vx_j}}{\int d^3\vx_i \, d^3\vx_j} = \frac{ (2\pi)^6 \delta_D^{(3)}(\vk) \delta_D^{(3)}(\vk') }{V^2}
\end{equation}
We have also assumed that there are no correlations between $\vx_i$ and $\vx_j$ (\textit{i.e.} no other source of clustering besides random chance). The power spectrum expression becomes
\begin{equation}
\begin{split}
    (2\pi)^3 \delta_D^{(3)}(\vk-\vk') P_\BH(k) & =
    (2\pi)^6 \delta_D^{(3)}(\vk) \delta_D^{(3)}(\vk') +
    \frac{1}{\bar{n}_\BH^2} \left( \sum_{i \neq j} \frac{ (2\pi)^6 \delta_D^{(3)}(\vk) \delta_D^{(3)}(\vk') }{V^2} + \sum_i \frac{ (2\pi)^3 \delta_D^{(3)}(\vk-\vk')}{V} \right)\\
    & ~~~~~~~~ - \frac{1}{\bar{n}_\BH} \left(  \sum_i \frac{(2\pi)^3\delta_D^{(3)}(\vk)}{V} (2\pi)^3\delta_D^{(3)}(\vk') +  \sum_j \frac{(2\pi)^3\delta_D^{(3)}(\vk')}{V} (2\pi)^3\delta_D^{(3)}(\vk) \right)
\end{split}
\end{equation}
The sums over $i$ yield a factor of the total number of black holes in the volume, $N$, and the sum over $i \neq j$ gives $N^2$ in the large $N$ limit:
\begin{equation}
\begin{split}
    \delta_D^{(3)}(\vk-\vk') P_\BH(k) &= \frac{1}{\bar{n}_\BH^2} \left( \frac{N^2}{V^2} (2\pi)^3 \delta_D^{(3)}(\vk) \delta_D^{(3)}(\vk') + \frac{N}{V} \delta_D^{(3)}(\vk-\vk') \right)\\
    & - \frac{2}{\bar{n}_\BH} \frac{N}{V} (2\pi)^3 \delta_D^{(3)}(\vk) \delta_D^{(3)}(\vk') + (2\pi)^3 \delta_D^{(3)}(\vk) \delta_D^{(3)}(\vk')
\end{split}
\end{equation}

Finally we can identify the ratio $N/V$ as the average black hole number density $\bar{n}_\BH$, and the zero-frequency terms cancel, leaving behind:
\begin{equation}
    \delta_D^{(3)}(\vk-\vk') P_\BH(k) = \frac{1}{\bar{n}_\BH} \delta_D^{(3)}(\vk-\vk') ~~\implies~~
    P_\BH(k) = \frac{1}{\bar{n}_\BH}
\end{equation}

\end{document}